\documentclass[]{aa}
\usepackage{times}
\usepackage{graphics}
\usepackage{epsf}

\begin{document}
 
   \thesaurus{09          % A&A Section 9: The Sun
              (02.03.3;   % Convection
               02.08.3;   % Hydrodynamics
               02.12.1;   % Line: formation
               02.18.7;   % Radiative transfer
               06.07.2;   % Sun: granulation
               06.16.2)}   % Sun: photosphere

%   \title{Solar Fe line formation with 3D hydrodynamical model atmospheres}
%   \subtitle{I. Shapes, shifts and asymmetries}
   \title{Line formation in solar granulation}
   \subtitle{I. Fe line shapes, shifts and asymmetries}

   \author{M. Asplund\inst{1,2}, \AA. Nordlund\inst{3},
           R. Trampedach\inst{4}, C. Allende Prieto\inst{5}, and
           R.F. Stein\inst{4}
          }

   \offprints{martin@astro.uu.se}

   \institute{
              NORDITA, 
              Blegdamsvej 17, 
              DK-2100 ~Copenhagen {\O}, 
              Denmark         
              \and
              present address: Uppsala Astronomical Observatory,
              Box 515,
              SE-751 20 ~Uppsala,
              Sweden
              \and
              Astronomical Observatory, NBIfAFG, 
              Juliane Maries Vej 30,
              DK-2100 ~Copenhagen \O, 
              Denmark
              \and
%              Theoretical Astrophysics Center, 
%              Ny Munkegade bldg. 520, 
%              DK-8000 \AA rhus C., 
%              Denmark\\
%              \and
              Dept. of Physics and Astronomy, 
              Michigan State University, 
              East Lansing, MI 48823, USA
              \and
              McDonald Observatory and Department of Astronomy,
              University of Texas,
              Austin, TX 78712-1083,
              USA 
              }

   \date{Received: January 24, 2000; accepted: May 4, 2000}

\authorrunning{M. Asplund et al.}
\titlerunning{Solar line formation: I. Fe line shapes, shifts
and asymmetries}

   \maketitle

%%%%%%%%%%%%%%%%%%%%%%%%%%%%%%%%%%%%%%%%%%%%%%%%%%%%%%%%%%%%%%%%%%%%
   \begin{abstract}

Realistic ab-initio 3D, radiative-hydrodynamical convection
simulations of the solar granulation have been applied to 
Fe\,{\sc i} and Fe\,{\sc ii} line formation.
% in order to 
%investigate the realism of such simulations in terms of spectral synthesis.
In contrast to classical analyses based on hydrostatic 1D model atmospheres
the procedure contains no adjustable free parameters but the treatment of the
numerical viscosity in the construction of the 3D, time-dependent,
inhomogeneous model atmosphere and the elemental abundance in the 3D spectral
synthesis. However, the numerical viscosity is introduced purely for
numerical stability purposes and is determined from standard hydrodynamical
test cases with no adjustments allowed to improve the agreement with
the observational constraints from the solar granulation.

The non-thermal line broadening is mainly provided by the Doppler shifts
arising from the convective flows in the solar photosphere and the
solar oscillations. 
The almost perfect agreement between the predicted temporally and 
spatially averaged line profiles for weak Fe lines with the observed profiles
and the absence of trends in derived abundances with line strengths,
seem to imply that the micro- and macroturbulence concepts are obsolete in 
these 3D analyses. 
Furthermore, the theoretical line asymmetries and shifts
%, which are caused by the convective overshoot motion and temperature-velocity 
%correlations in the line-forming regions, 
show a very satisfactory
agreement with observations with an accuracy of typically
$50-100$\,m\,s$^{-1}$ on an absolute velocity scale.
%The intermediate strong ($W_\lambda \sim 10$\,pm) 
%Fe lines are, however, not perfectly described with a tendency for 
%slightly too deep and too much redshifted cores.
The remaining minor discrepancies point to how the convection 
simulations can be refined further. 
%most notably in terms of numerical resolution and 
%frequency-dependent radiative transfer with allowance for
%Doppler shifts. 

   \end{abstract}

%%%%%%%%%%%%%%%%%%%%%%%%%%%%%%%%%%%%%%%%%%%%%%%%%%%%%%%%%%%%%%%%%%%%
\section{Introduction}
%%%%%%%%%%%%%%%%%%%%%%%%%%%%%%%%%%%%%%%%%%%%%%%%%%%%%%%%%%%%%%%%%%%%

A proper understanding of convection has for a long time
been one of the greatest challenges for stellar astrophysics.
Since convection can strongly influence both the stellar evolution and
the emergent stellar 
spectra, this uncertainty also carries over to other fields of
astrophysics, such as galactic evolution and cosmology.  
Convection also plays a 
dominant role in various processes besides energy transport
and stellar evolution, perhaps most notably 
mixing of nuclear-processed material, 
magnetic activity, chromospheric heating
and stellar pulsations. 
Convection is furthermore of relevance in 
searches for extra-solar planets by introducing differential spectral line
shifts that are likely to change periodically with solar-like cycles,
and may then affect the measurements of periodic variations induced by
orbiting companions.
%typical differential line shifts are on the order of 500\,m\,s$^{-1}$
%for stars such as the Sun (e.g. Dravins et al. 1981), while the reflex
%motion due to the gravitational tug is on the order of 10\,m\,s$^{-1}$
%for Jupiter-like exo-planets.
%The coupling between convection and pulsation is still
%largely unexplored but may well affect the use of e.g. Cepheid stars
%for measuring cosmic distances. 
%{\bf General feature in astrophysical objects (e.g. accretion disks)?}

The convection zone in late-type stars extends 
to the visible surface layers and thereby 
influences the emergent spectrum, from which 
virtually all information about the stars is deduced. 
For the Sun, the directly observable manifestation is granulation: 
an evolving pattern of brighter (warmer) rising material with 
darker (cooler) sinking gas in between, 
driven by the radiative cooling in a thin surface layer. 
Convection also leaves distinct signatures in the spectral
lines in the form of line shifts and asymmetries
(e.g. Gray 1980; Dravins et al. 1981; Dravins 1982;
Dravins \& Nordlund 1990a,b).
For late-type stars, the bisectors of unblended lines have
a characteristic $\subset$-shape, which results from the differences
in line strength, continuum level, area coverage and Doppler
shift between up- (granules) and downflows (intergranular
lanes). Weak lines normally only show the upper part of
the $\subset$-shape and have convective blueshifts (after removing
the effects of the gravitational redshift). The cores of
stronger lines are formed in or above the convective overshoot
layers and therefore have smaller or non-existent blueshifts
(Allende Prieto \& Garc\'{\i}a L{\'o}pez 1998a,b).
Solar transition region and coronal emission lines can, however, 
show significant 
redshifts of several km\,s$^{-1}$ (Doschek et al. 1976). 

%The field of observing line asymmetries and interpreting those
%in terms of the physical conditions in the upper part of
%the convection zone was pioneered by D.F. Gray and D. Dravins
%with collaborators (e.g. Gray 1980; Dravins et al. 1981). 
The possibility to obtain spectra of the Sun on
an absolute wavelength scale (i.e. corrected for the radial velocity
of the Sun relative to the spectrograph) 
with exceptionally high spectral resolution, signal-to-noise ratio (S/N)
and wavelength coverage,
makes the Sun an ideal testbench to study the physical conditions
in the upper parts of the convection zone.
For other stars one has previously been restricted to only a few spectral
lines and poorer quality observations.
Fortunately the situation has recently improved significantly
(e.g. Allende Prieto et al. 1999, 2000),
most notably concerning resolution, S/N and the use of carefully
wavelength-calibrated spectra covering a large portion of the
visual region. 

In order to draw the correct conclusions from the observed
stellar spectrum it is necessary to have a proper
understanding of how convection modifies the atmospheric
structure and the spectral line formation. 
Up to the present, theoretical models of stellar atmospheres 
have been based on several unrealistic assumptions,
such as restriction to a 1-dimensional (1D)
stratification in hydrostatic equilibrium with only a 
rudimentary parametrisation of convection through e.g. the mixing
length theory (B\"ohm-Vitense 1958) or some close relative thereof
(Canuto \& Mazzitelli 1991). 
Such models therefore miss entirely the very nature of convection
as an inhomogeneous, non-local and time-dependent phenomenon,
which is driven by the radiative cooling at the surface.
Furthermore, spectral synthesis using 
classical 1D model atmospheres show poor resemblance with
observed spectra. In particular, the predicted line broadening
is much too weak (of course, with no knowledge about the convective
motions and Doppler shifts, all lines are also perfectly symmetric).
To partly cure the problem, two artificial fudge parameters are introduced
which the theoretical spectrum is convolved with to mimic the missing
broadening: micro- and macroturbulence.
In this situation, systematic errors arising from modeling inadequacies
in e.g. chemical analyses of stars are far
larger than the measuring errors alone.
An improved theoretical foundation for the interpretation
of stellar spectra is therefore highly desirable, to 
complement the impressive recent observational advances
in terms of S/N, spectral resolution and 
limiting magnitudes following the new generation of large-aperture telescopes,
like the ESO VLT and McDonald HET, and detectors. 

With the advent of modern supercomputers, self-consistent 
3D radiative hydrodynamical simulations of
the surface convection in stars have now become tractable
(e.g. Nordlund \& Dravins 1990; Stein \& Nordlund 1989, 1998;
Asplund et al. 1999; Trampedach et al. 1999).
%through the pioneering work of \AA . Nordlund (Copenhagen).
%Sofar such detailed modelling has only been applied to the Sun
%and a few similar stars. 
%It is noteworthy that t
These simulations invoke no free parameters like mixing length parameters
except the numerical viscosity for stability purposes,
%in sharp contrast to classical model atmospheres, 
yet succeed in reproducing key observational diagnostics such as
granulation topology and statistics (Stein \& Nordlund 1989, 1998), and
helioseismic properties (Rosenthal et al. 1999).
%and spectral
%line shapes and asymmetries (Dravins et al. 1981; Dravins \& Nordlund 1990).
In the present paper we present a detailed study of the spectral 
line formation of Fe lines using such solar surface convection 
simulations as 3D model atmospheres. The resulting line shapes,
shifts and asymmetries are found to compare very satisfactory
with the observations, lending strong support to the realism 
of the simulations. 
The derived solar Fe abundances from weak and strong Fe\,{\sc i}
and Fe\,{\sc ii} lines are presented in an accompanying article
(Asplund et al. 2000b, hereafter Paper II). Subsequent papers in
the series will deal with Si lines (Asplund 2000, hereafter Paper III)
and observed and predicted 
spatially resolved solar lines (Kiselman \& Asplund, in preparation, 
hereafter Paper IV). Additionally, hydrogen line formation will
be the topic of a following article (Asplund et al., in preparation,
hereafter Paper V).

%%%%%%%%%%%%%%%%%%%%%%%%%%%%%%%%%%%%%%%%%%%%%%%%%%%%%%%%%%%%%%%%%%%%
\section{Surface convection simulations}
%%%%%%%%%%%%%%%%%%%%%%%%%%%%%%%%%%%%%%%%%%%%%%%%%%%%%%%%%%%%%%%%%%%%

The 3D model atmospheres of the solar granulation which form
the basis for the spectral line calculations presented here have
been obtained with a 3D, time-dependent, compressible, 
radiative-hydrodynamics code developed to study solar and stellar
surface convection (e.g. Nordlund \& Stein 1990; 
Stein \& Nordlund 1989, 1998; Asplund et al. 1999). 
The hydrodynamical equations of
mass, momentum and energy conservation:

\begin{equation}
\frac{\partial {\rm ln} \rho}{\partial t} = - {\bf \bar{v}} \cdot \nabla 
{\rm ln} \rho - \nabla \cdot {\bf \bar{v}}
\end{equation}

\begin{equation}
\frac{\partial {\bf \bar{v}}}{\partial t} = - {\bf \bar{v}} \cdot \nabla 
{\bf \bar{v}} + {\bf \bar{g}}
- \frac{P}{\rho} \nabla {\rm ln} P + \frac{1}{\rho} \nabla \cdot \sigma
\end{equation}

\begin{equation}
\frac{\partial e}{\partial t} = - {\bf \bar{v}} \cdot \nabla e 
- \frac{P}{\rho} \nabla \cdot {\bf \bar{v}} + Q_{\rm rad} + Q_{\rm visc}
\end{equation}

\noindent coupled to the equation
of radiative transfer along a ray (in total eight inclined rays):

\begin{equation}
{\rm d}I_\lambda/{\rm d}\tau_\lambda = I_\lambda - S_\lambda
\end{equation}

\noindent are solved on a non-staggered Eulerian mesh with 
200\,x\,200\,x\,82 gridpoints; simulations with
resolutions 100\,x\,100\,x\,82, 50\,x\,50\,x\,82 and
50\,x\,50\,x\,41
have also been computed but have not been utilized here since they
are hampered by less good agreement between predicted and observed
line profiles (Asplund et al. 2000a).
%(Asplund, Ludwig \& Nordlund, in preparation).
In these equations, $\rho$ denotes the density,
${\bf \bar{v}}$ the velocity, ${\bf \bar{g}}$ the gravitational acceleration,
$P$ the pressure, $e$ the internal energy,
$\sigma$ the viscous stress tensor, $Q_{\rm visc}$ the viscous dissipation,
$Q_{\rm rad} = \int_\lambda \int_\Omega \kappa_\lambda 
(I_\lambda-S_\lambda){\rm d}\Omega{\rm d}\lambda$ 
the radiative heating/cooling rate, $I_\lambda$ the monochromatic intensity,
$S_\lambda = B_\lambda(T)$ the corresponding source function,
$\tau_\lambda$ the optical depth and $\kappa_\lambda$ the absorption
coefficient.

The physical dimension of the simulation box corresponds to 
6.0\,x\,6.0\,x\,3.8\,Mm of which about 1.0\,Mm is located above 
continuum optical depth unity. Again, initial solar simulations
extending only to heigths of 0.6\,Mm suffered from problems in
the predicted line asymmetries for strong lines which was traced
to the influence of the outer boundary and that non-negligible optical
depths were present in the line cores already at the outermost layers.
Though not completely removed with the current more extended
simulations, the problems have largely disappeared as discussed
further in Sect. \ref{s:shift} and \ref{s:asym}. 
The depth scale has been optimized
to provide the best resolution where it is most needed, i.e. 
in those layers with the steepest
gradients in terms of d$T$/dz and d$^2T$/dz$^2$, which for the Sun
occurs around the visible surface (which is defined to have a geometrical
depth $z = 0$\,Mm). Through this procedure, sufficient 
depth coverage was also obtained in the important optically thin 
layers. The simulation box covers about 13 pressure scale heights 
and 11 density scale heights in
depth, while the horizontal dimension is sufficient to include $\ga 10$
granules at any time of the simulation. 
The spatial derivatives are computed using third order splines
and the time-integration is a third-order leapfrog predictor-corrector
scheme (Hyman 1979; Nordlund \& Stein 1990). The code was stabilized
using a hyper-viscosity diffusion algorithm (Stein \& Nordlund 1998) with
the parameters determined from standard hydrodynamical test cases, like
the shock tube. In this respect these parameters are not freely
adjustable parameters for the surface convection simulations. 
It is important to realize that changing the numerical
resolution in effect also varies the effective viscosity, which shows that
the convective efficiency and temperature structure are independent on
the adopted viscosity description (Asplund et al. 2000a).
Periodic horizontal boundary conditions and open transmitting top
and bottom boundaries were used in an identical fashion to
the simulations described by Stein \& Nordlund (1998). 
The influence of the boundary conditions on the results are negligible.
In particular the upper boundary has been placed at sufficiently great
heights not to disturb the line formation process except for the cores
of very strong lines, and the lower boundary is located at large 
depths to ensure that the inflowing gas is isentropic and featureless.
No magnetic fields or rotation were included
in the present calculation.
%(though preliminary MHD simulations have recently been 
%performed by Stein \& Nordlund (1999) at lower numerical resolution). 

In order to obtain a realistic atmospheric structure it is crucial
to correctly describe the internal energy of the gas and the
energy exchange between radiation and gas. For this purpose 
a state-of-the-art equation-of-state (Mihalas et al. 1988), which
includes the effects of ionization, excitation and dissociation,
has been used. 
Since the solar photosphere is located in the
layers where the convective energy flux from the interior is
transferred to radiation, a proper treatment of the 3D radiative
transfer is necessary, which has been included under the 
assumption of LTE and using detailed continuous (Gustafsson et al. 1975
and subsequent updates) and line (Kurucz 1993) opacities. 
The 3D equation of radiative transfer was solved at each timestep of the
convection simulation for eight inclined rays (2 $\mu$-angles and
4 $\varphi$-angles) using the opacity binning technique (Nordlund 1982). 
The four opacity bins correspond to continuum, weak lines, intermediate
strong lines and strong lines; shorter tests with eight instead of
four opacity bins revealed only very minor differences in the resulting
atmospheric structures. The accuracy of the binning procedure was verified
throughout the simulation at regular intervals 
by solving the full monochromatic
radiative transfer (2748 wavelength points) in the 1.5D approximation 
(i.e. treating each vertical column separately in the flux calculations
without considering the influence from neighboring columns)
and found to always agree within 1\% in emergent flux. 

It is noteworthy that the 
convection simulations contain no adjustable free parameters besides those
used to characterize the stars: the effective temperature $T_{\rm eff}$
(or equivalently, as adopted here, the entropy of the inflowing 
material at the bottom boundary), 
the surface acceleration of gravity log\,$g$ and the
chemical composition. The entropy at the lower boundary was carefully
adjusted prior to the simulation started 
in order to reproduce the total solar luminosity. The solar 
surface gravity (log\,$g=4.437$\,[cgs]) was employed. 
The chemical composition of the gas were
taken from Grevesse \& Sauval (1998), in particular the Fe abundance
used for the equation-of-state and opacity calculations was 
7.50\footnote{On the customary logarithmic abundance scale defined
to have log$\,\epsilon_{\rm H}=12.00$}. The line opacities were interpolated
to the adopted Fe abundance using the standard Kurucz ODFs with 
log$\,\epsilon_{\rm Fe}=7.67$ and non-standard ODFs computed with 
log$\,\epsilon_{\rm Fe}=7.51$ and without any He
(Kurucz 1997, private communication, cf. Trampedach 1997 for details
of the procedure).

%%%%%%%%%%%%%%%%%%%%%%%%%%%%%%%%%%%%%%%%%%%%%%%%%%%%%%%%%%%%%%%%%%%%
\section{Spectral line calculations and atomic data}
%%%%%%%%%%%%%%%%%%%%%%%%%%%%%%%%%%%%%%%%%%%%%%%%%%%%%%%%%%%%%%%%%%%%

From the full solar convection simulation, which covers two solar hours,
%(corresponding to roughly one CPU-month on a supercomputer
%such as the Fujitsu VX-1), 
a shorter sequence of 50\,min with snapshots stored every 30\,s
was chosen for the subsequent spectral line calculations.
The adopted snapshots have a temporally averaged effective temperature 
very close to the observed nominal $T_{\rm eff} = 5777\pm3\,$K for the Sun
(Willson \& Hudson 1988):
$5767 \pm 21$\,K where the quoted range is the standard deviation 
of the individual snapshots. 
For our present purposes the time coverage is sufficient to 
obtain statistically significant 
spatially and temporally averaged line profiles and shifts,
as verified by test calculations 
after dividing the snapshots in various sub-groups.
Even as short sequences as 10\,min produced line asymmetries and
shifts to within 50\,m\,s$^{-1}$ of the full calculations.
In order to improve the vertical
sampling, the simulation was interpolated to a finer depth scale 
with the same number of depth points but only extending 
down to layers with minimum continuum optical depths of 
$\tau_{\rm cont} = 300$ ($\approx 700$\,km below $\tau_{\rm cont} = 1$
rather than $2.9\,$Mm in the original simulation).
Prior to the line transfer computations, the snapshots were also 
interpolated to a coarser grid of 50\,x\,50\,x\,82 but with the same
horizontal dimension to save computing time; various tests assured
that the procedure did not introduce any differences in the spatially
averaged line shapes or asymmetries. 
The Doppler shifts introduced by the convective motions in the
3D model atmosphere were correctly accounted for in the solution
of the 3D spectral line transfer. 
In most cases, intensity profiles at the center of the 
solar disk ($\mu = 1.0$) were considered,
which have been calculated for every column of the interpolated 
snapshots, before spatial and temporal averaging and normalization.
A few lines where, however, computed under different viewing angles to
allow a disk-integration for comparison with published solar flux atlases
(Kurucz et al. 1984).
The assumption
of LTE in the ionization and excitation balances and for the
source function in the line transfer calculations ($S_\nu = B_\nu$)
have been made throughout in the present study. 
The background continuous opacities were calculated using
the Uppsala opacity package (Gustafsson et al. 1975 and subsequent updates).

Fe lines are the most natural tools to study line shapes and
asymmetries in stars: Fe is an abundant element with a complex
atomic structure which ensures there
are many useful lines of the appropriate strength; there exists
accurate laboratory measurements for the necessary atomic data
such as transition probabilities and wavelengths (e.g. Blackwell et al. 1995;
Nave et al. 1994);
the Fe nuclei have a large atomic mass which minimizes the thermal
velocities; the influence from isotope and hyperfine splitting should be
negligible (e.g. Kurucz 1993); and LTE is a reasonable approximation for Fe
(e.g. Shchukina \& Trujillo Bueno 2000), at least 
for 1D model atmospheres of solar-type stars 
(3D NLTE studies of Fe may, however, reveal
larger departures from NLTE as speculated by e.g. Nordlund 1985 and 
Kostik et al. 1996). By studying a large sample of neutral and 
ionized Fe lines with different atomic data and therefore line
strengths, the solar photospheric convection at varying 
atmospheric layers can be probed by
analysing the resulting line shifts and asymmetries.
The Fe lines and their atomic data are the same as described in detail in 
Paper II. In particular, accurate laboratory wavelengths for the Fe\,{\sc i}
and Fe\,{\sc ii} lines were taken from Nave et al. (1994) and 
Johansson (1998, private communication), respectively. 
The final profiles have been computed with the individual 
Fe abundances derived in Paper II. 

For comparison with observations, the solar FTS disk-center intensity
atlas by Brault \& Neckel (1987) (see also Neckel 1999) 
has been used, due to its superior
quality over the older Liege atlas by Delbouille et al. (1973) 
in terms of wavelength calibration 
(Allende Prieto \& Garc\'{\i}a L{\'o}pez 1998a,b) and continuum tracement.
%for very strong lines (the Balmer lines are e.g. noticably asymmetric in
%the Liege atlas).
For flux profiles the solar atlas by Kurucz et al. (1984) has been used,
which is also based on FTS-spectra with a similar spectral resolution
as the disk-center atlas. 
The wavelengths for the observed profiles have been adjusted to remove
the effects of the solar gravitational redshift (633\,m\,s$^{-1}$).
All spatially averaged theoretical profiles have
been convolved with an instrumental profile 
to account for the finite spectral resolution 
%(resolving power $R = $ at XXX\,nm) 
of the observed atlas.
Since the atlas was acquired with a Fourier Transform Spectrograph
(FTS), the instrumental profile corresponds to a sinc-function 
% Kurucz flux atlas: UV: R=350000, IR: R=520000, lambda>480nm: R=520000
% lambda: 300-1300nm
% Brault & Neckel intensity atlas: R~350000, lambda=330-1250nm
with $\lambda/\Delta\lambda \simeq 520\,000$ in the visual
rather than the normal Gaussian (e.g. Gray 1992).
The additional instrumental broadening has only a minor (but not entirely
negligible) effect on the
resulting line asymmetries due to the high resolving power of the FTS.

%%%%%%%%%%%%%%%%%%%%%%%%%%%%%%%%%%%%%%%%%%%%%%%%%%%%%%%%%%%%%%%%%%%%
\section{General features of 3D line formation}
%%%%%%%%%%%%%%%%%%%%%%%%%%%%%%%%%%%%%%%%%%%%%%%%%%%%%%%%%%%%%%%%%%%%

The convective motions and the atmospheric 
inhomogeneities leave distinct fingerprints in the spectral
lines, which can be used 
%when observed at very high spectral resolution and S/N 
to decipher the conditions
in the line-forming layers produced by the convection
(e.g. Dravins et al. 1981; Dravins \& Nordlund 1990a,b). 
%These signatures take the form of line
%asymmetries and line shifts, which for the Sun are characterized
%by $\subset$-shape bisectors and convective blueshifts (after
%removing the effects of the gravitational redshift) when considering
%spatially averaged profiles. 
Line strengths of weak lines are predominantly determined by the average 
atmospheric temperature structure, while the line widths reflect the
amplitude of the Doppler shifts introduced by the velocity fields. 
Line shifts and bisectors are created by the correlations between 
temperature and velocity and the details of the convective overshooting,
as well as the statistical distribution between up- and downflows.
%Due to the large variation of individual profiles across the granulation
%pattern, the average profile and its departure from
%perfect symmetry are sensitively dependent on the atmospheric
%inhomogeneities and velocities. 
Therefore, obtaining a good 
agreement between observed and predicted profiles lend strong support
to the realism of the simulations.  

%%%%%%%%%%%%%%%%%%%%%%%%%%%%%%%%%%%%%%%%%%%%%%%%%%%%%%%%%%%%%%%%%%%%
\subsection{Spatially resolved profiles  \label{s:prof_xy}}
%%%%%%%%%%%%%%%%%%%%%%%%%%%%%%%%%%%%%%%%%%%%%%%%%%%%%%%%%%%%%%%%%%%%

\begin{figure}[t]
\resizebox{\hsize}{!}{\includegraphics{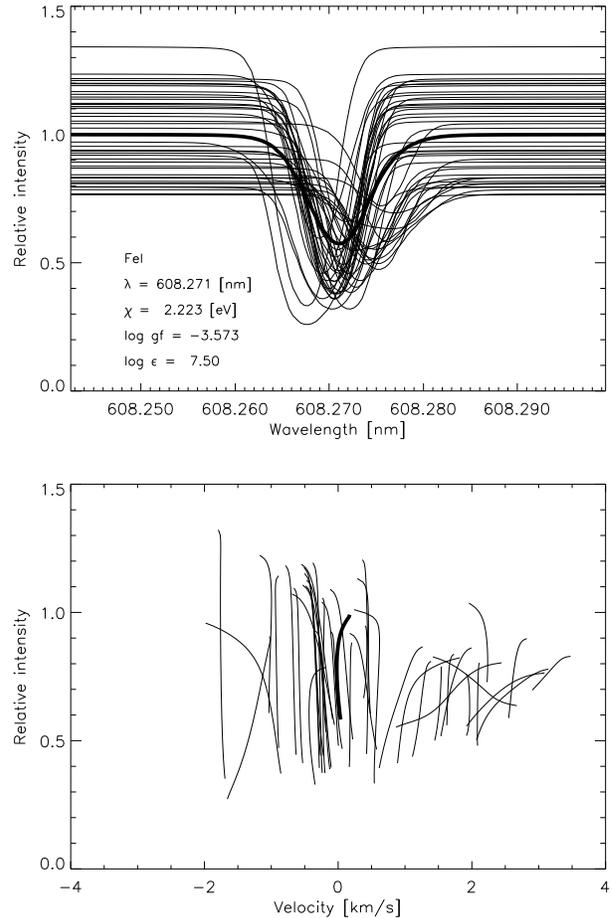}}
%\resizebox{\hsize}{!}{\includegraphics{figures/FeI6082_xy_bw2.ps}}
%\centerline{
%\psfig{figure=figures/FeI6082_xy_bw.ps,width=9.cm}}
%  \picplace{1cm}
\caption{Spatially resolved profiles and bisectors for the 
Fe\,{\sc i} 608.2 line. The lines are both stronger and have
higher continuum intensities in the blueshifted granules 
compared with the red-shifted intergranular lanes. 
The largest vertical velocities in the photosphere are encountered
in the downflows.
The intensity scale is normalized to the spatially averaged
continuum level. The thick solid lines correspond to the
spatially averaged profile and bisector
}
         \label{f:prof_xy}
\end{figure}

%\begin{figure}[t]
%\resizebox{\hsize}{!}{\includegraphics{figures/FeI6082_W.ps}}
%\caption{Spatially resolved equivalent widths for the 
%Fe\,{\sc i} 608.2 line as a function of the local 
%relative continuum intensity ({\it upper panel})
%and vertical velocity at the depth with average
%continuum optical depth unity ({\it lower panel});
%negative velocities correspond to upflows. 
%The lines are in general both stronger and have
%higher continuum intensities in the granules 
%compared with the intergranular lanes. The equivalent widths
%have been computed assuming an iron abundance of 7.50.
%The horizontal dashed lines represent the spatially averaged
%equivalent width. No atmospheric or instrumental smearing have 
%been applied to the theoretical profiles
%}
%         \label{f:w_xy}
%\end{figure}

Although Paper IV and V in the present series of articles will discuss
in detail observed and predicted spatially resolved lines, a brief
excursion is still warranted here in order to interpret the
spatially averaged profiles and bisectors presented in
Sects. \ref{s:shape}, \ref{s:shift} and \ref{s:asym}.

%The temperature variations over the granulation naturally result
%in intensity variations between granules and intergranular lanes.
%The continuum intensity brightness contrast of the raw solar 
%simulation is $I_{\rm rms}^{\rm cont}/<I^{\rm cont}> \simeq 16$\% 
%(Fig. \ref{f:prof_xy}) at 610\,nm.
%When convolving the intensities with a suitable point-spread-function
%representing the atmospheric and instrumental seeing, the intensity
%contrast decreases to the best observed value of about 8\%
%(Lites et al. 1989) 
%The comparison between observed and simulated intensity contrasts
%is, however, severely complicated by poorly understood 
%details of the adopted instrumental
%and atmospheric point-spread functions (Nordlund 1984).
%Due to the temperature dependence of the opacity,
%$I_{\rm rms}$ is smaller than the temperature contrast
%in the continuum forming layers which have $T_{\rm rms}/<T> \simeq 24$\%: 
%the surface with continuum optical depth unity is corrugated and thus the 
%high temperature gas is partly hidden from sight
%(Stein \& Nordlund 1998). 

Spatially resolved profiles 
%carry information which can be used to infer the
%different atmospheric structures in the up- and 
%downflows. The individual profiles 
take on an astonishing range of shapes and shifts spanning
several km\,s$^{-1}$, as illustrated in Fig. \ref{f:prof_xy}.
The intensity contrast reversal in the higher layers of the photosphere,
i.e. granules become dark while intergranular lanes become bright a few
hundred km above the visible surface, is clearly seen in the cores of
the resolved profiles. 
%As illustrated in Fig. \ref{f:w_xy}, 
The strengths of spatially averaged profiles 
are normally biased towards
the granules, since the upflows in general are brighter (high
continuum intensity), have steeper temperature gradients and
have a larger area coverage than the downflows. These trends are
also observationally confirmed,
which suggests that the combination of 3D hydrodynamical model atmospheres and
LTE is appropriate for most Fe lines 
(Kiselman 1998; Kiselman \& Asplund 2000; Paper IV). 
%This bias towards granules may
%explain the fact that different mixing length parameters $\alpha_{\rm MLT}$
%seem to be required for 1D spectral synthesis and evolutionary 
%calculations (Fuhrmann et al. 1994): 
%the line formation favours the granular structures
%which have steeper temperature gradients ($\alpha_{\rm MLT} \approx 0.5$)
%while evolutionary models mainly depend on the interior entropy structure,
%which are characterized by $\alpha_{\rm MLT} \approx 1.6$ for the
%Sun (Ludwig et al. 1999; Trampedach et al. 1999).
%%We emphasize, however, that a priori it is not possible to assume
%%that a 1D model with a specific $\alpha_{\rm MLT}$, whether 0.5 or 1.5,
%%is appropriate for all purposes, even all 
%%spectroscopic features; the same is of course true for models
%%based on the Canuto \& Mazzitelli (1991) recipe
%%(Gardiner et al. 1999). It would be naive
%%to hope that any 1D model atmosphere can satisfactory contain all the
%%details of the 3D velocity field and inhomogeneities, though it {\em may}
%%work well for some spectral features. In this context it can be noted that
%%Grevesse \& Sauval (1998, 1999) are forced to adjust the 
%%temperature structure of the Holweger-M\"uller
%%(1974) model differently for C, N, O and Fe lines to remove existing
%%trends with excitation potential. 
%%No such ad-hoc procedures are necessary
%%with 3D model atmospheres (Paper II).

Spatially resolved bisectors are not at all typical of
the spatially averaged bisectors, which are merely the result of
the statistical distribution of individual profiles. Rather than 
the characteristic $\subset$-shape bisectors, individual bisectors
normally have inverted $\backslash$-shape bisectors
in granules and $\slash$-shapes in intergranular lanes, as seen
in Fig. \ref{f:prof_xy}. 
This reflects the in general increasing vertical velocities with
depth in the photosphere. However, due to the meandering motion
of the downflows, occasionally the line-of-sight passes through both
upward and downward moving material which causes large
variations for certain columns. 
%This effect is even more obvious
%in spectra obtained closer to the limb (Dravins \& Nordlund 1990a).
%The bisector variation shown in Fig. \ref{f:prof_xy} is 
%unfortunately significantly larger than what can be observed today
%due to the smearing by instrumental and atmospheric seeing
%(cf. Dravins et al. 1981; Steffen \& Freytag 1991), but
%recent progress in terms of telescope design and 
%adaptive optics will certainly improve the observational situation 
%(e.g. Scharmer et al. 1999).

%%%%%%%%%%%%%%%%%%%%%%%%%%%%%%%%%%%%%%%%%%%%%%%%%%%%%%%%%%%%%%%%%%%%
\subsection{Spatially averaged disk-center profiles}
%%%%%%%%%%%%%%%%%%%%%%%%%%%%%%%%%%%%%%%%%%%%%%%%%%%%%%%%%%%%%%%%%%%%

\begin{figure}[t]
\resizebox{\hsize}{!}{\includegraphics{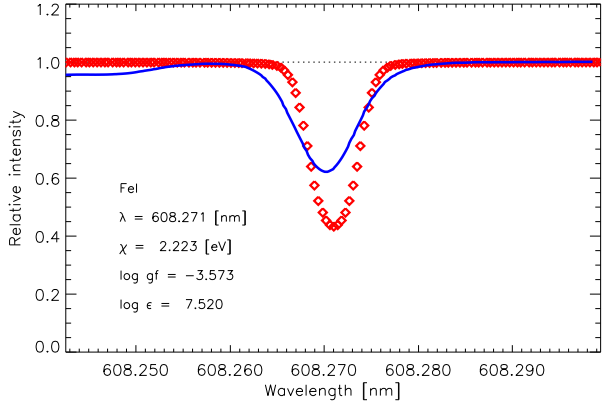}}
%\resizebox{\hsize}{!}{\includegraphics{figures/FeI6082.7_v=0.ps}}
%\centerline{
%\psfig{figure=figures/FeI6082.7_v=0.ps,width=9.cm}}
%  \picplace{1cm}
\caption{The spatially and temporally averaged Fe\,{\sc i}
608.2 line when artificially removing all convective velocities
in the simulations (diamonds). In comparison with the solar
intensity atlas (solid line, Brault \& Neckel 1987) the predicted line
is much too narrow and lacks the correct line shift and asymmetry.
The result when including the self-consistent velocity field is
shown in Fig. \ref{f:prof}
}
         \label{f:prof_v=0}
\end{figure}

\begin{figure}[t]
\resizebox{\hsize}{!}{\includegraphics{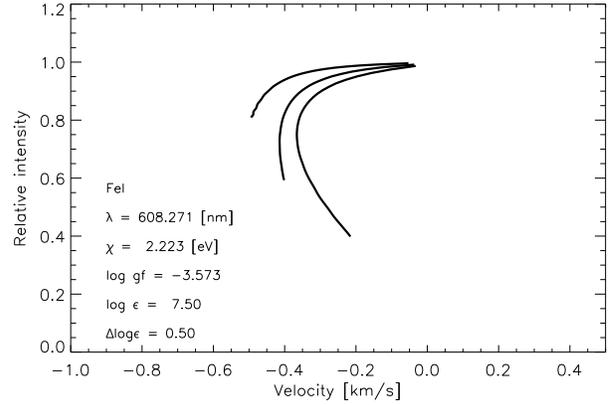}}
%\resizebox{\hsize}{!}{\includegraphics{figures/FeI6082_bis_allabund.ps}}
%\centerline{
%\psfig{figure=figures/FeI6082_bis_allabund.ps,width=9.cm}}
%  \picplace{1cm}
\caption{The spatially and temporally averaged Fe\,{\sc i}
608.2 bisector assuming three different abundances: 
log\,$\epsilon_{\rm Fe} = 7.00$, 7.50 and 8.00 (or equivalently with
three different log\,$gf$-values). Note that the upper parts
of the bisectors do not coincide for the different 
line strengths since the whole region of line formation is
shifted outwards for stronger lines
%, i.e. the parts of the bisectors
%with a specific line depth do not necessarily originate in the
%same atmospheric layers
}
         \label{f:bis_abund}
\end{figure}

\begin{figure}[t]
\resizebox{\hsize}{!}{\includegraphics{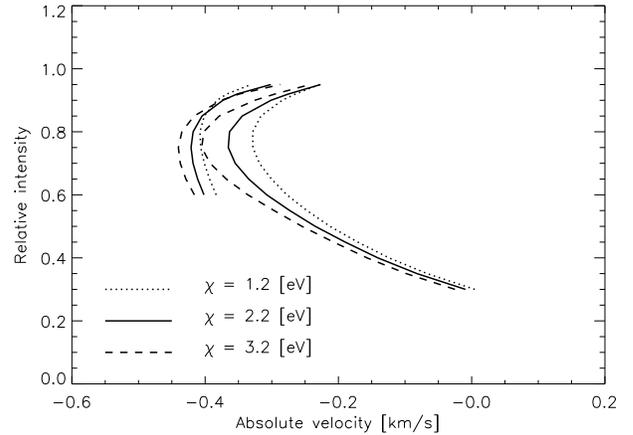}}
%\resizebox{\hsize}{!}{\includegraphics{figures/FeI_bis_chi_bw.ps}}
%\centerline{
%\psfig{figure=figures/FeI_bis_chi.ps,width=9.cm}}
%  \picplace{1cm}
\caption{The bisectors of the Fe\,{\sc i} 608.2 (weak) 
and 621.9\,nm (strong) lines (solid lines) which both 
have $\chi_{\rm exc} = 2.2\,$eV. Also shown are the corresponding (artificial)
Fe\,{\sc i} lines with $\chi = 1.2\,$eV (dotted lines) and $\chi = 3.2\,$eV
(dashed lines). The fake lines have all other transition
properties the same as the original lines but with the $gf$-values
($\simeq \pm 1.0\,$dex) adjusted to return the same line depths.
All lines have here been computed for the same 5\,min sequence
from the full convection simulation, which means that the 
bisectors differ slightly from the average of the whole 50\,min simulation
%used in the comparison with the observed lines in 
%Sect. \ref{s:shape}, \ref{s:shift} and \ref{s:asym}.
%With a lower excitation potential the line formation region
%is shifted outwards which causes smaller convective blueshifts
}
         \label{f:bis_chi}
\end{figure}

\begin{figure}[t]
\resizebox{\hsize}{!}{\includegraphics{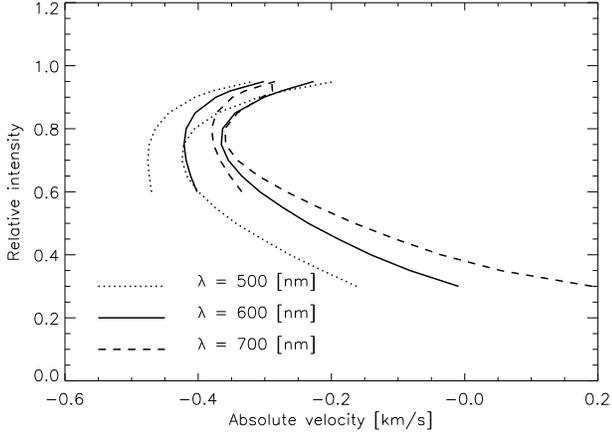}}
%\resizebox{\hsize}{!}{\includegraphics{figures/FeI_bis_lam_bw.ps}}
%\centerline{
%\psfig{figure=figures/FeI_bis_lam.ps,width=9.cm}}
%  \picplace{1cm}
\caption{The bisectors of the Fe\,{\sc i} 608.2 (weak) 
and 621.9\,nm (strong) lines (solid lines).
Also shown are the corresponding (artificial) Fe\,{\sc i} 
lines with $\lambda = 500$\,nm (dotted lines) and $\lambda = 700$\,nm
(dashed lines). 
%Yes, I know, the 500nm line should of course be blue, 600nm green and 
%700nm red
The fake lines have all other transition
properties the same as the original lines but with the $gf$-values
adjusted to return the same line depths. 
All lines have here been computed for the same 5\,min sequence
as in Fig. \ref{f:bis_chi}
%. The increased brightness contrast and
%steeper temperature gradients for shorter wavelengths tend to 
%increase the convective blueshifts when more vigorous convective motions
%are seen in the deeper layers
}
         \label{f:bis_lam}
\end{figure}

The widths of spatially averaged spectral lines, which clearly exceed the
natural and thermal broadening, predominantly arise from the 
velocity amplitude of the granules and intergranular lanes and
to a lesser extent from photospheric oscillations. 
%With a proper treatment of the atmospheric velocity field there appears to be
%no need to invoke the concepts of micro- and macroturbulence as illustrated
%below and in Paper II and III,
%which are required in spectral synthesis using classical 1D model atmospheres. 
A demonstration of the importance of the non-thermal
Doppler broadening is presented
in Fig. \ref{f:prof_v=0}, which shows the resulting spatially
averaged profile from the 3D simulations but with 
all convective velocities artificially set equal to zero in the
line calculation; thereby
the predicted profile closely resemble those calculated with classical
1D model atmospheres. Clearly, without the Doppler shifts 
the line is much too narrow, which requires additional broadening
in the form of micro- and macroturbulence to be introduced.
%To partly cure this limitation in 1D one has to resort to introduce 
%which are supposed to represent the missing turbulent motions on 
%length scales less than (micro-) and larger than (macroturbulence) 
%a unit optical path length. Such a simplistic division is naturally
%artificial since the motions occur on a range of scales in the
%photosphere. These extra broadening parameters (as well as 
%the shapes of the convolution profile, e.g. a Gaussian or
%a radial-tangential distribution, Gray 1992) are then adjusted
%to achieve the correct line width for a given line strength. 
%However, in spite of having 
%several free parameters the detailed line shapes are not properly
%described, which reflects the intrinsic limitations of the procedure
%and the 1D models (Fig. \ref{f:prof}). 
The poor agreement between observations and predictions when not
including the self-consistent velocity field as shown in 
Fig. \ref{f:prof_v=0}, should be
contrasted with the excellent fit shown in Fig. \ref{f:prof}.

The individual line bisectors depends on the details of the
line formation and thus on transition properties such as 
log\,$gf$, $\chi_{\rm exc}$ and $\lambda$ in an intricate
way, as illustrated in Fig. \ref{f:bis_abund}, \ref{f:bis_chi} and 
\ref{f:bis_lam}. When discussing mean bisectors 
(e.g. Gray 1992; Allende Prieto et al. 1999; Hamilton \& Lester 1999) 
it is therefore important to
consider only appropriate subgroups consisting of lines with
similar characteristics to avoid introducing errors when interpreting the
results in terms of convective properties. 
As a corollary, it
follows that reconstructing a mean bisector by averaging bisectors or
using shifts of lines of different
strengths will in general not recover in 
detail the individual bisectors, as exemplified in Fig. \ref{f:bis_abund}.
%Such a mean bisector will be biased towards 
%smaller velocity spans compared
%with the individual bisectors for weak lines in stars with 
%characteristic $\subset$-shape bisectors. 
%On the other hand, a bisector derived from 
%the shifts of lines of various line depths 
%(Hamilton \& Lester 1999) will also be distorted compared with individual
%bisectors, producing too much blueshift for stronger lines
%in the upper parts of the bisectors. 
As expected, for line depths $\ga 0.5$ 
the bisectors closely coincide for lines of different strengths
due to the disappearance of
the influence from the convective inhomogeneities.
Similarly, decreasing the excitation potential tends to shift the
line formation outwards, producing less convective blueshifts
for a given line depth
(Fig. \ref{f:bis_chi}), while decreasing the wavelength increases
the brightness contrast and makes the temperature gradients steeper,
resulting in more vigorous convective line asymmetries (Fig. \ref{f:bis_lam}).
%The difficulty in observationally confirm the dependence on excitation
%potential (Hamilton \& Lester 1999) is clearly related to the problem
%of finding suitable lines with otherwise the same properties. 
%This illustrates the power of combining
%observational line asymmetry studies with 3D hydrodynamical modeling,
%which allows individual lines to be considered.

\begin{figure}[t]
\resizebox{\hsize}{!}{\includegraphics{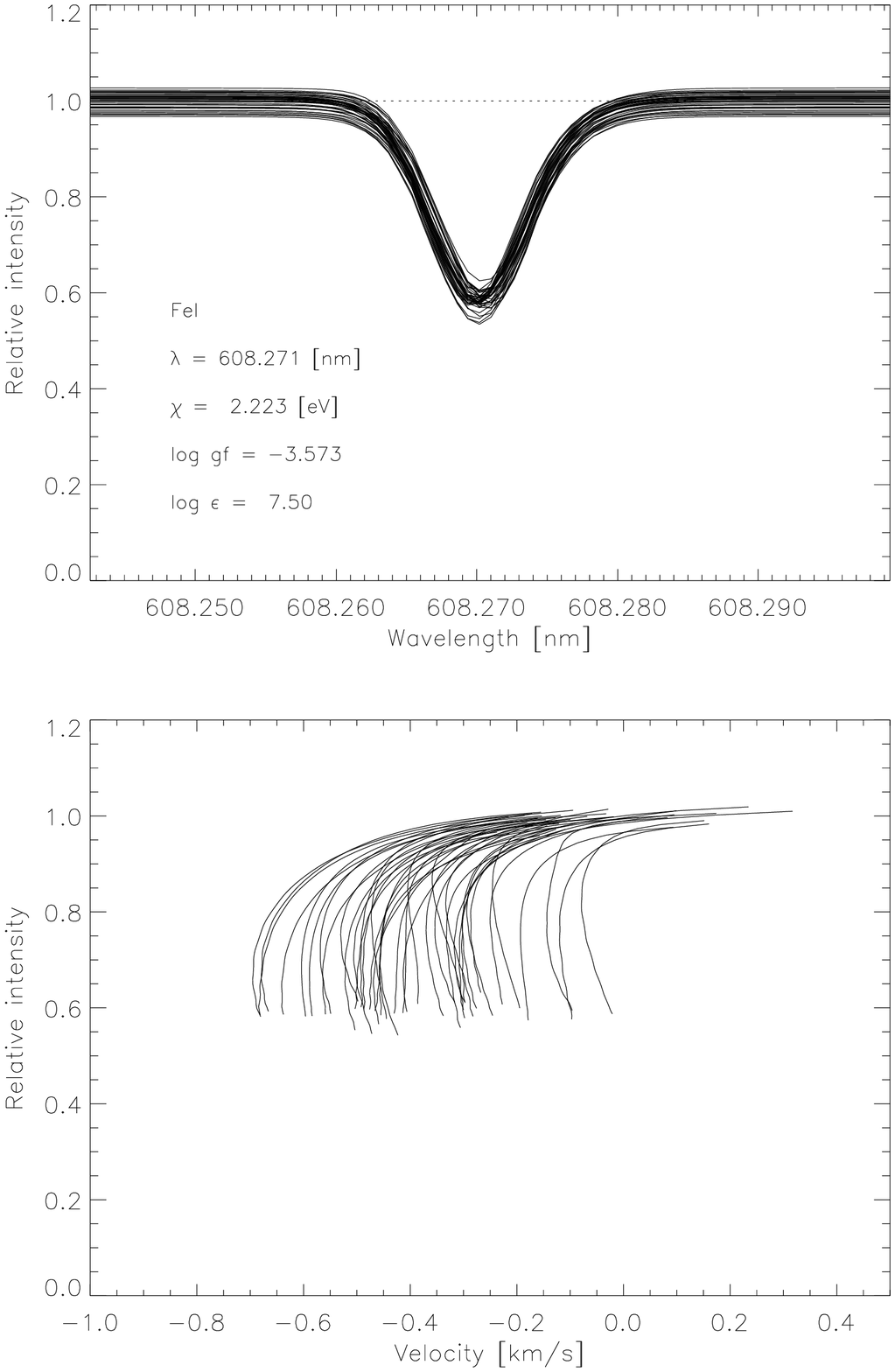}}
%\resizebox{\hsize}{!}{\includegraphics{figures/FeI6082_t_bw.ps}}
%\centerline{
%\psfig{figure=figures/FeI6082_t.ps,width=9.cm}}
%  \picplace{1cm}
\caption{The temporal evolution of the spatially 
averaged profile and bisector for the 
Fe\,{\sc i} 608.2 line. The 40 profiles shown here are taken
at one minute intervals from the full solar simulation. 
The intensity scale is normalized to the averaged
continuum level for the full simulation sequence
%Clearly seen in the bisectors are the oscillations present
%in the numerical box, which corresponds to the solar
%5\,min oscillations
}
         \label{f:prof_t}
\end{figure}

The temporal evolution of the granulation pattern 
also introduce changes in the line profiles.
Solar granules have typical lifetimes of about 10\,min but 
since the numerical
box covers typically $\ga 10$ granules at any time, the influence of
growing and decaying granules are relatively modest, as illustrated
in Fig. \ref{f:prof_t}.
%; in 2D the temporal variations are much
%more pronounced since roughly 100 times fewer gridpoints are used
%and therefore correspondingly longer time sequences
%are required to obtain statistically significant line profiles
%(Asplund et al. 2000b).
%(Asplund, Ludwig \& Nordlund, in preparation).
Since the emergent $T_{\rm eff}$ is not enforced but rather is the
result of the evolving granulation pattern, the continuum level varies
slightly ($\approx \pm 2\%$, corresponding to $T_{\rm eff, rms}=20\,$K)
throughout the simulation, as well as the line strength. 
The bisector shapes are only slightly modified by the granulation, though
the presence of oscillations in the simulation box 
shifts the bisectors back and forth in a regular fashion with a period
corresponding to the solar 5\,min oscillations. 
These numerical oscillations have an amplitude
of about $\pm 300$\,m\,s$^{-1}$ in the line-forming layers.
The oscillations additionally broaden the line, 
though without altering the line
strength since essentially the whole photosphere oscillates in unison
without modifying the atmospheric structure.
%; in this sense the 
%oscillations work in a similar way as macroturbulence for 1D spectral
%synthesis.

%%%%%%%%%%%%%%%%%%%%%%%%%%%%%%%%%%%%%%%%%%%%%%%%%%%%%%%%%%%%%%%%%%%%
\subsection{Spatially averaged off-center intensity and 
disk-integrated flux profiles}
%%%%%%%%%%%%%%%%%%%%%%%%%%%%%%%%%%%%%%%%%%%%%%%%%%%%%%%%%%%%%%%%%%%%

Although not the main emphasis of the present paper a few 
Fe lines have been computed at different viewing angles
(4 $\mu$-angles and 4 $\varphi$-angles) in
order to enable a disk-integration to obtain flux profiles.
A disk-integration requires further that the rotational broadening 
(1.8\,km\,s$^{-1}$ in the case of the Sun) is taken into account.
The line formation of flux profiles is more complex than for
intensity profiles due to the contributions from different disk positions.
The well-known limb-darkening decreases the continuum intensities
towards the limb but the lines also tend to be weaker due to the
more shallow temperature gradient in the upper atmosphere. 
Furthermore, the granulation contrast decreases towards the limb, since
higher layers with progressively smaller influence of the granulation
is seen. At a given instant the line asymmetries vary significantly more 
towards the limb as the line-of-sight may pass through both granules and
intergranular lanes and also the horizontal velocities may introduce
Doppler shifts. Thus the Doppler shifts are less well correlated with
the background continuum intensity, which introduces a more random nature
of the resulting bisectors for inclined line-of-sights.

\begin{figure}[t]
\resizebox{\hsize}{!}{\includegraphics{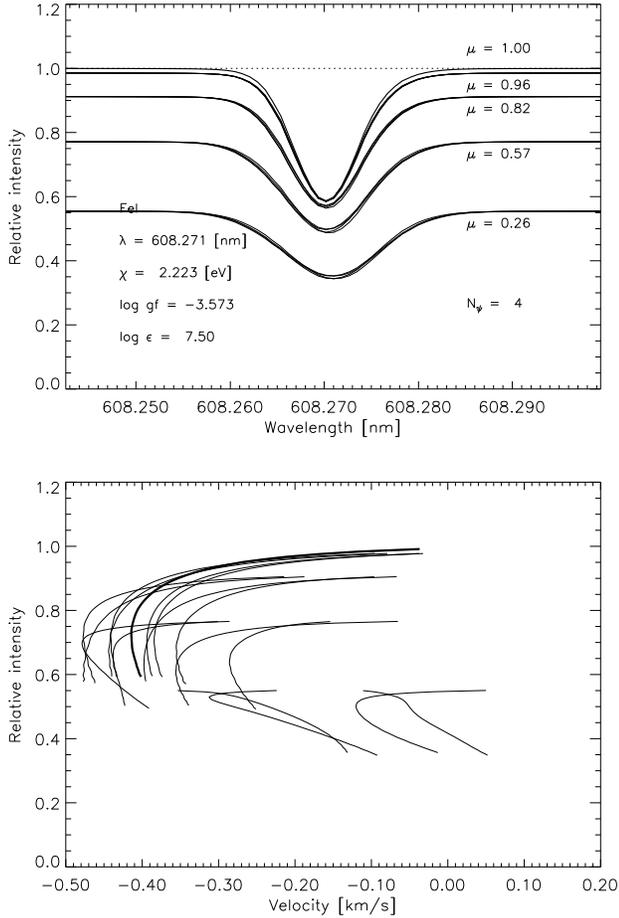}}
%\resizebox{\hsize}{!}{\includegraphics{figures/FeI6082_mu.ps}}
\caption{The spatially and temporally averaged 
Fe\,{\sc i} 608.2 profile and bisector at different
viewing angles. The line has been computed at 4 $\mu$-angles and
4 $\varphi$-angles. The profiles have been normalized to the
continuum intensity at disk center. The thick solid lines represent
the profile and bisector of the line at disk-center. Note the
"limb effect" in the bisectors: smaller convective blue-shift toward 
the limb
}
         \label{f:prof_mu}
\end{figure}

Fig. \ref{f:prof_mu} illustrates the variation of the spatially 
and temporally averaged intensity profiles and
bisectors of a weak Fe\,{\sc i} line at various viewing angles, which
make up the necessary profiles for a disk-integration. Also seen
is the limb-effect (Halm 1907): the wavelength of solar lines increases
towards the limb such that at small $\mu$ the typical convective
blueshift seen at disk-center has disappeared completely or even
been reversed into a small red-shift when removing the effects
of gravitational redshift and rotation. The limb-effect 
is the same effect as present between strong and weak lines, namely that
the core of the lines are formed in high enough photospheric layers where
the granulation contrast and velocities have largely vanished.
The red-shifted cores at very small $\mu$, which are both observed and
predicted with the 3D model atmospheres, are likely due to a bias for
receding horizontal velocities to be viewed against the higher
temperatures above intergranular lanes while the approaching gas will
preferentially be seen against the lower temperatures above granules
due to the temperature reversal in the higher photospheric layers
(Balthasar 1985).
%Note that our explanations differ from that suggested by Dravins (1982)
%who argued that the receding gas will be in front of bright granules.

%%%%%%%%%%%%%%%%%%%%%%%%%%%%%%%%%%%%%%%%%%%%%%%%%%%%%%%%%%%%%%%%%%%%
\section{Solar Fe line shapes \label{s:shape}}
%%%%%%%%%%%%%%%%%%%%%%%%%%%%%%%%%%%%%%%%%%%%%%%%%%%%%%%%%%%%%%%%%%%%

\begin{figure*}[t]
\resizebox{\hsize}{!}{\includegraphics{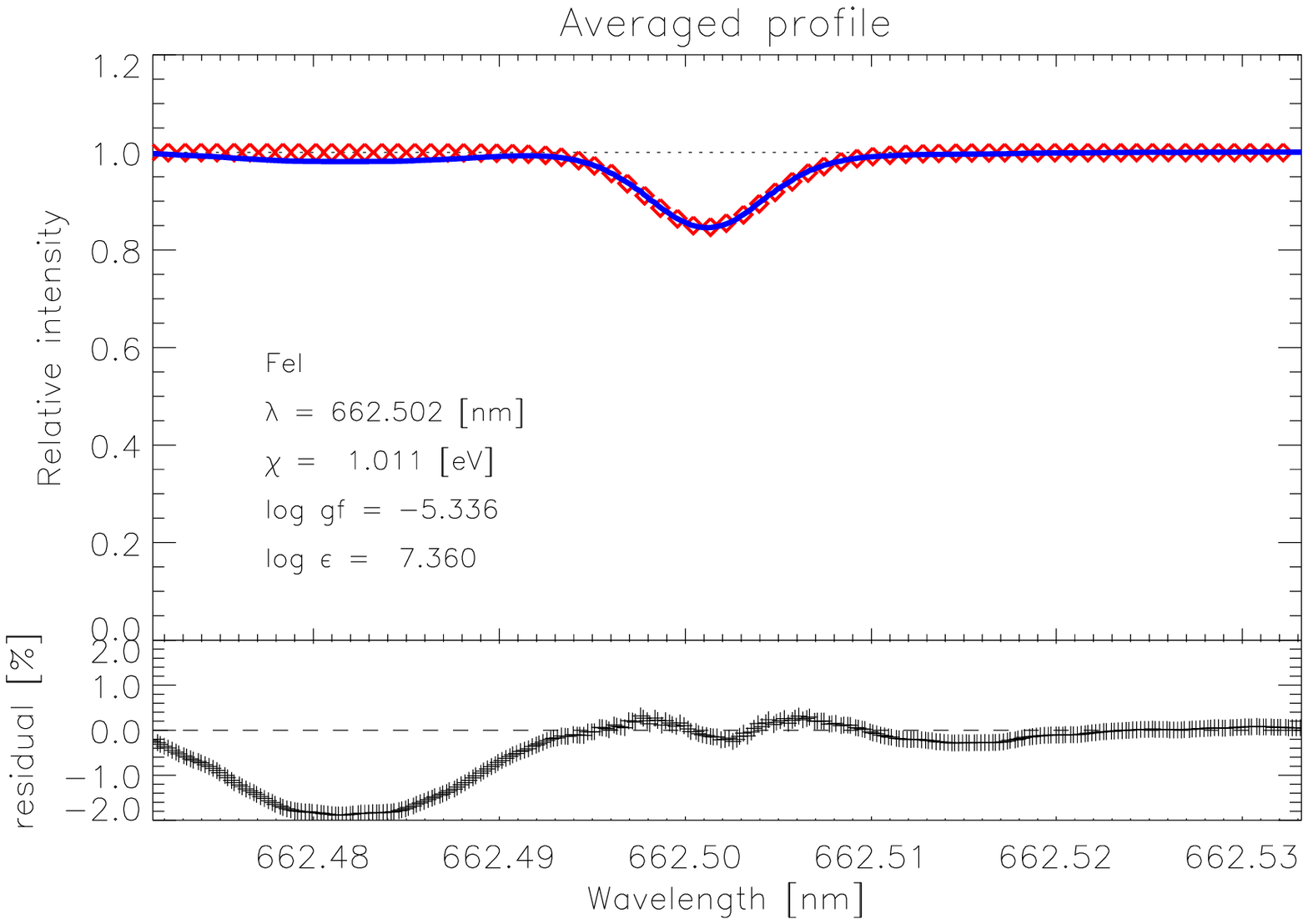}
\includegraphics{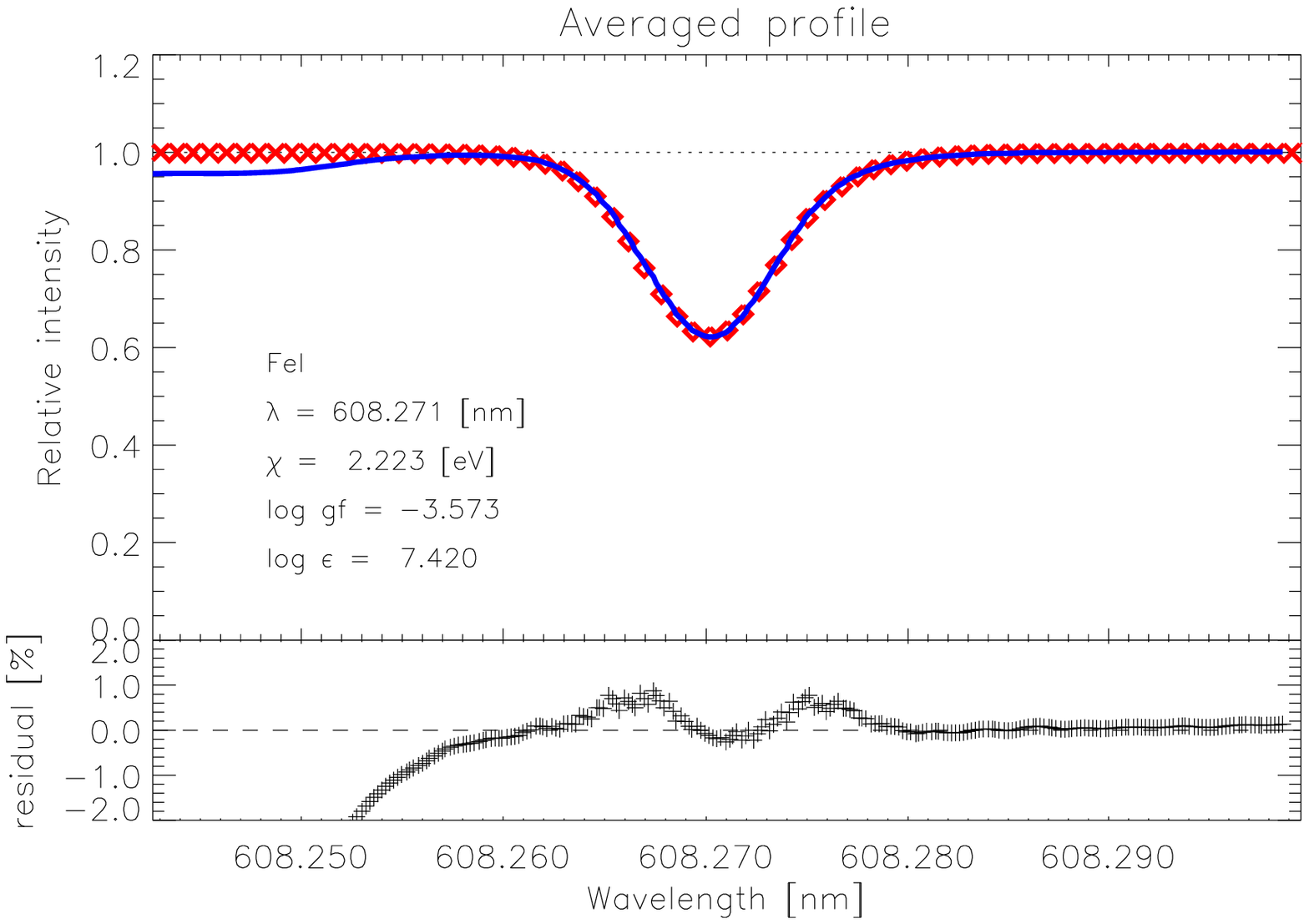}}
\resizebox{\hsize}{!}{\includegraphics{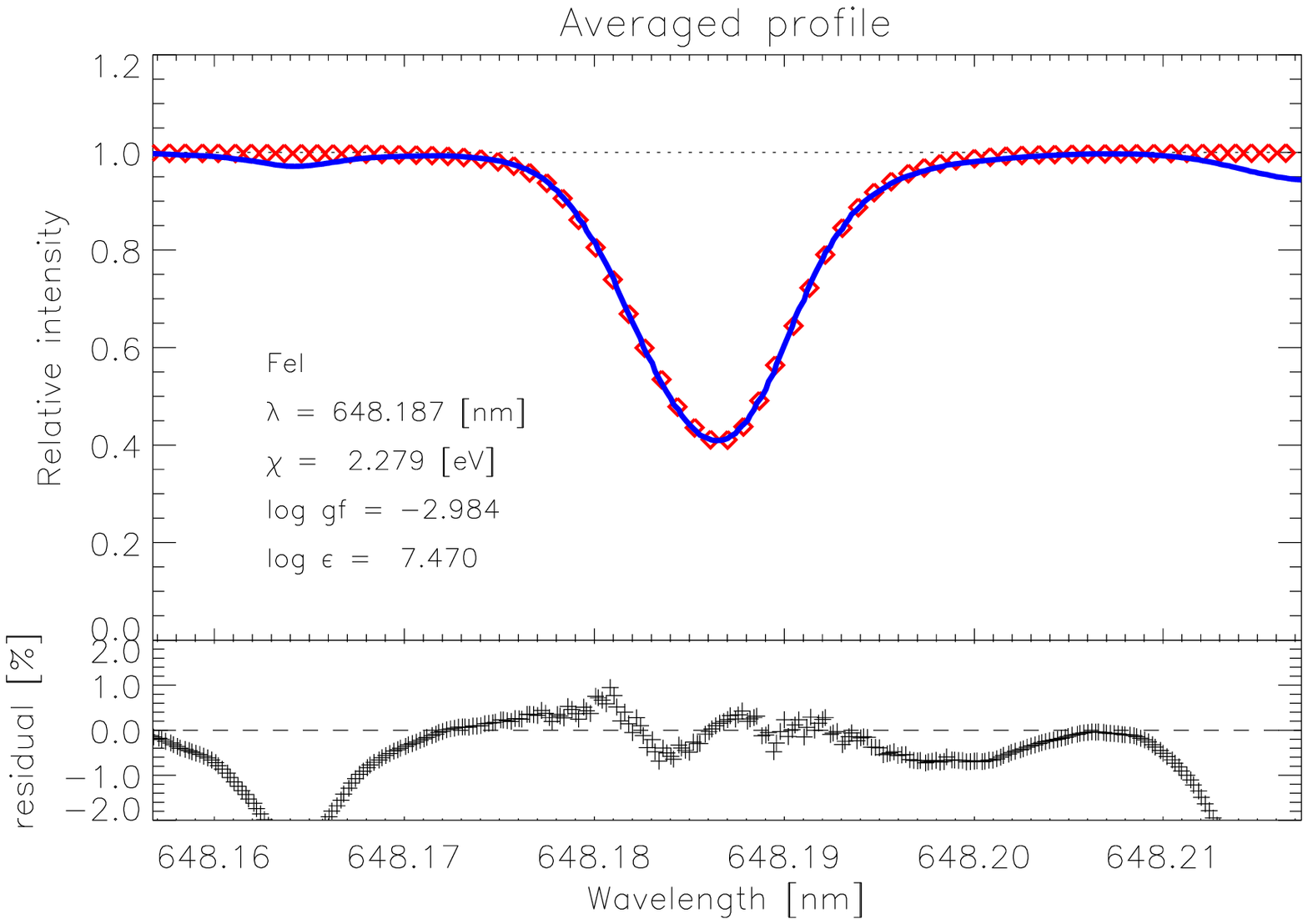}
\includegraphics{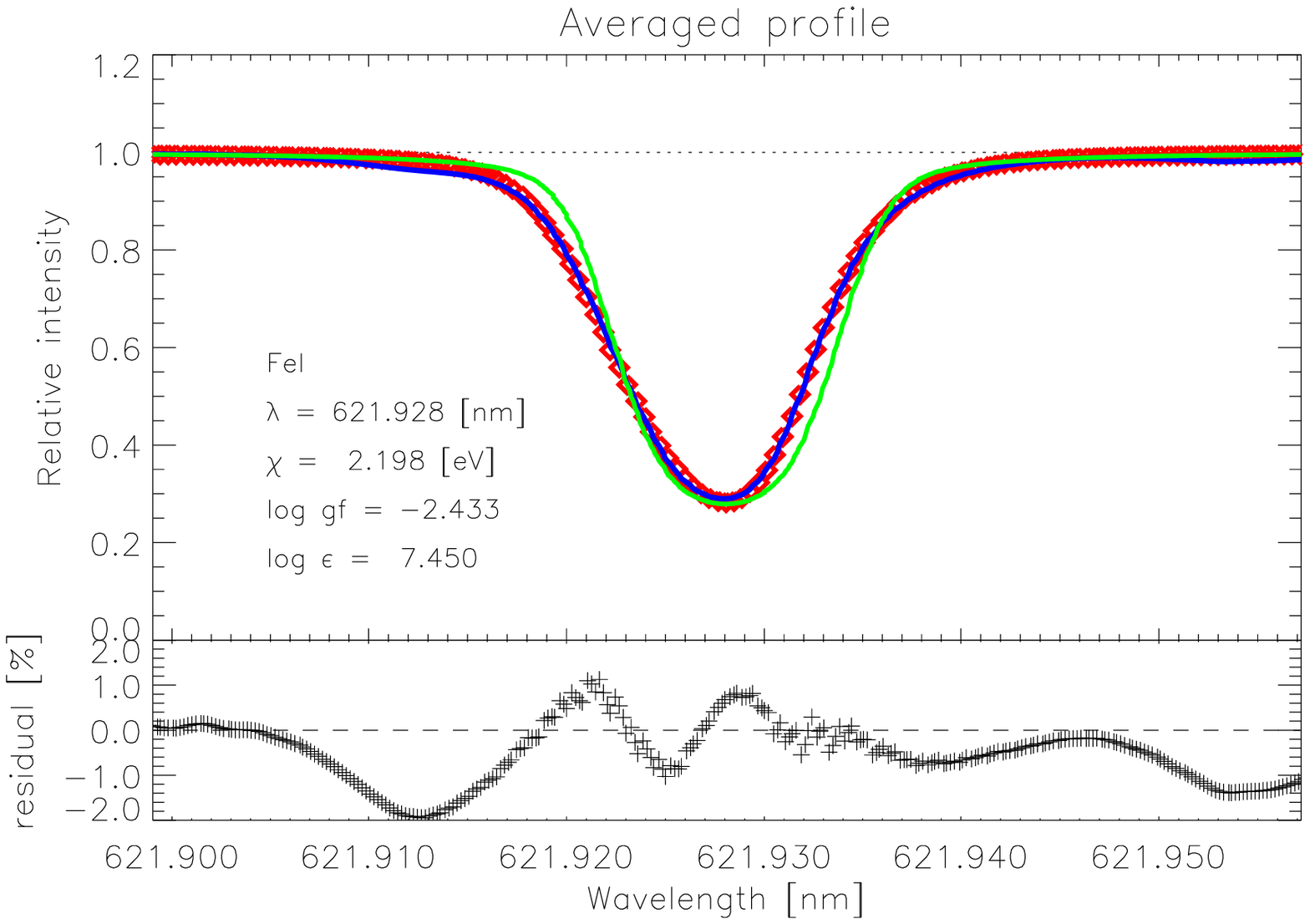}}
\resizebox{\hsize}{!}{\includegraphics{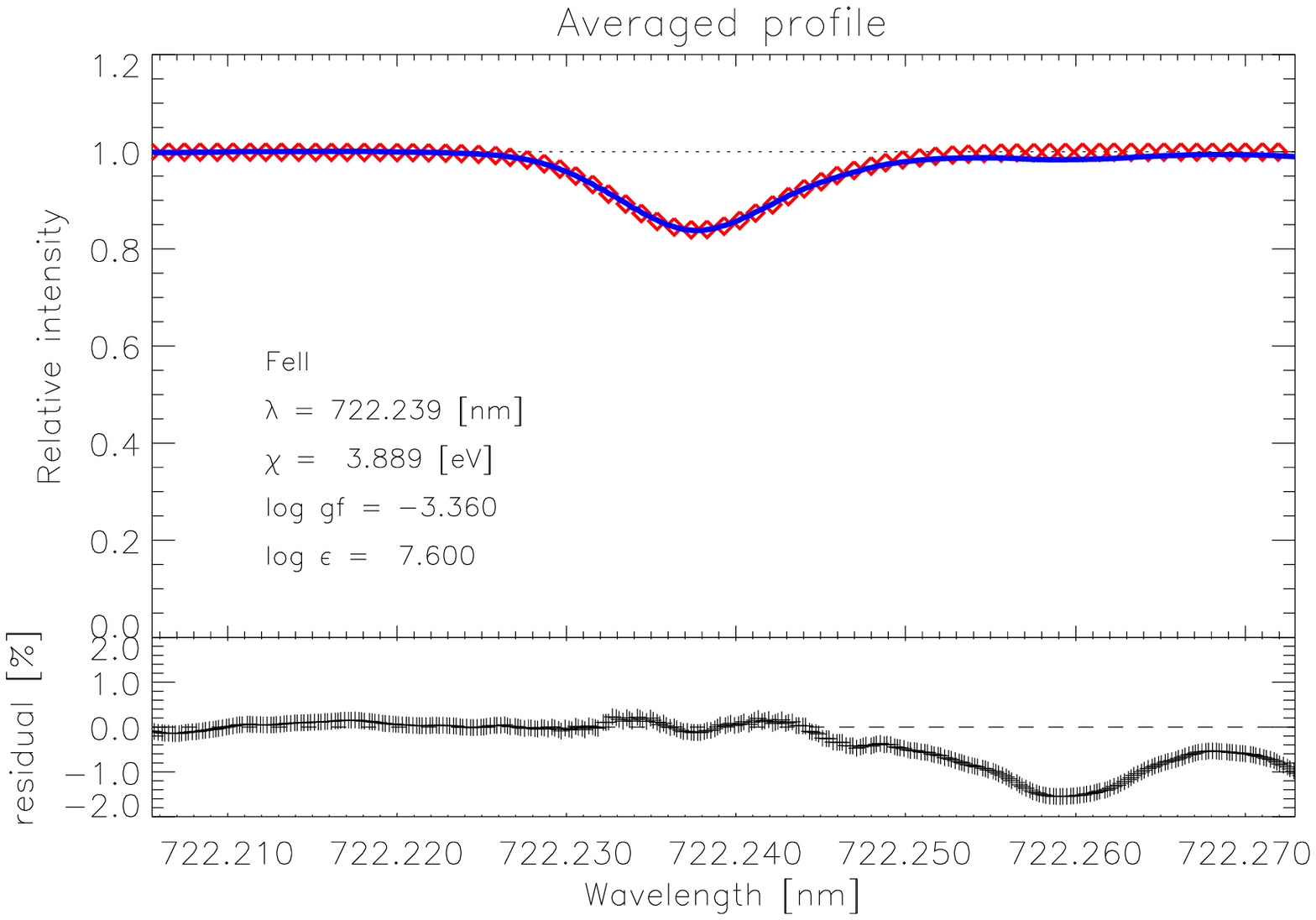}
\includegraphics{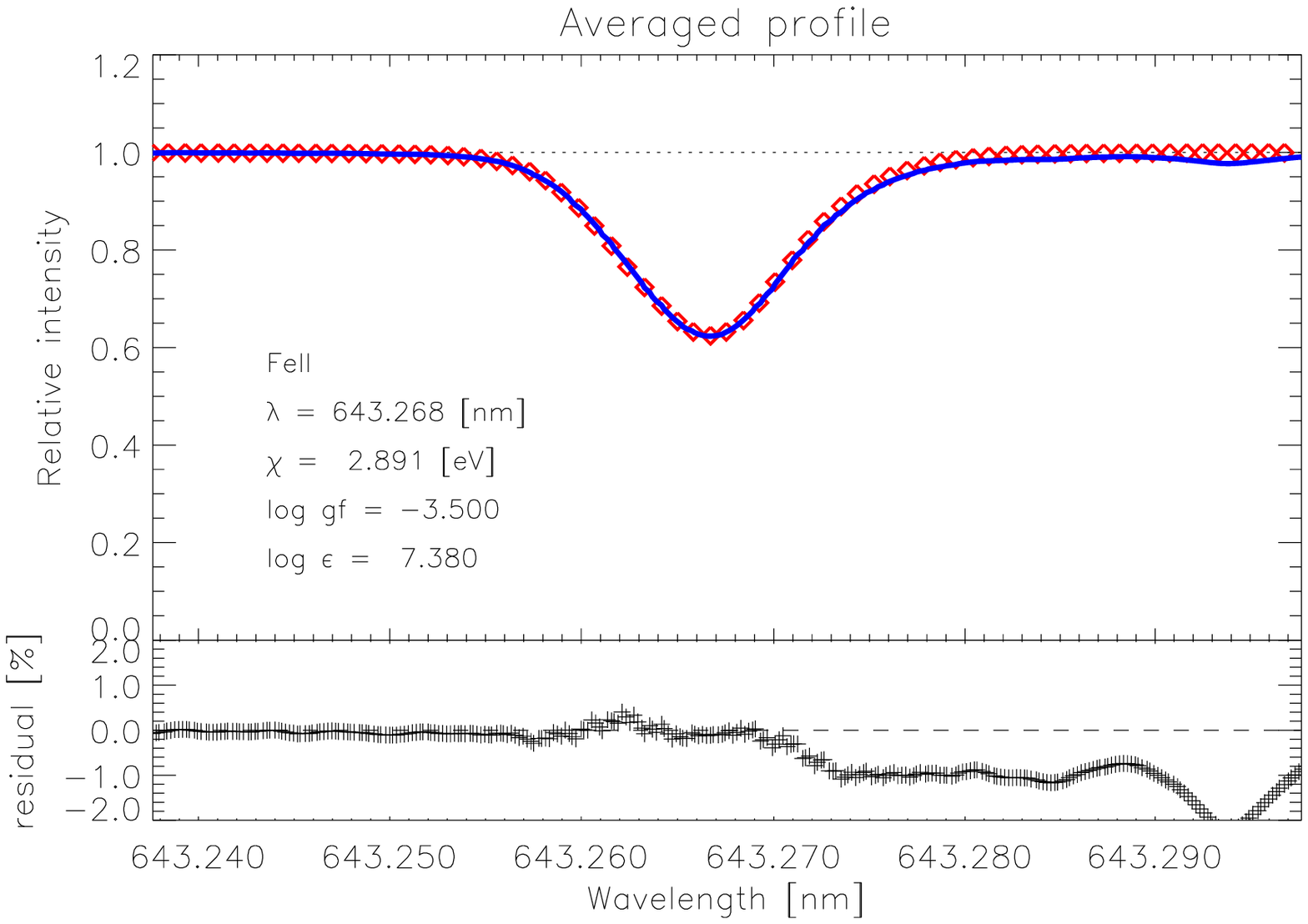}}
%\resizebox{\hsize}{!}{\includegraphics{figures/FeI6625.0_res.ps}
%\includegraphics{figures/FeI6082.7_res.ps}}
%\resizebox{\hsize}{!}{\includegraphics{figures/FeI6481.8_res.ps}
%\includegraphics{figures/FeI6219.2_res_1D.ps}}
%\resizebox{\hsize}{!}{\includegraphics{figures/FeII7222.3_res.ps}
%\includegraphics{figures/FeII6432.6_res.ps}}
\caption{Some examples of spatially and temporally averaged 
%weak (Fe\,{\sc i}\,608.7\,nm) and strong (Fe\,{\sc i}\,621.9\,nm)
disk-center Fe\,{\sc i} and Fe\,{\sc ii} 
lines (diamonds) compared with the observed solar atlas
(solid lines, Brault \& Neckel 1987). Also shown are the residual
intensities (observed - predicted) to emphasize the remaining minor
differences; minor blends not included in the spectral synthesis are
clearly seen in some of the line wings. 
The predicted profiles have been convolved
with a sinc-function to account for the finite spectral resolution
of the solar atlas. Minor corrections to the observed wavelengths
($<100$\,m\,s$^{-1}$) and continuum level ($<1\%$) have been allowed
to improve the fits. The Fe lines are almost perfectly matched
by the theoretical profiles, which implies that the rms velocity
amplitudes in the simulations are very close to the real values
in the solar photosphere. For comparison the best fit 1D profile
for the Fe\,{\sc i} 621.9\,nm line is also shown (green solid line
in the middle right panel), which has been computed with the Holweger-M\"uller
(1974) model atmosphere,
log$\,\epsilon_{\rm Fe I}=7.60$ (to achieve the same equivalent width
as the 3D profile), $\xi_{\rm turb} = 0.845$\,km\,s$^{-1}$
and convolved with a Gaussian macroturbulence of 1.6\,km\,s$^{-1}$ to
have the correct line depth
(radial-tangential macroturbulence is not applicable for
intensity profiles, Gray 1992); clearly the agreement is much
inferior in spite of the adjustable broadening parameters due to the
neglect of convective velocities and spatial inhomogeneities
%. Stronger lines are also well described
%although there is a tendency for a minor discrepancy in the
%line core, which is also reflected in the line asymmetries 
%(Sects. \ref{s:shift} and \ref{s:asym}). However, the cores of
%such strong lines are influenced by the 
%conditions in the higher atmospheric layers, which are likely the
%least realistic in the simulations given the inherent assumptions
%such as LTE
}
         \label{f:prof}
\end{figure*}

\begin{figure}[t]
\resizebox{\hsize}{!}{\includegraphics{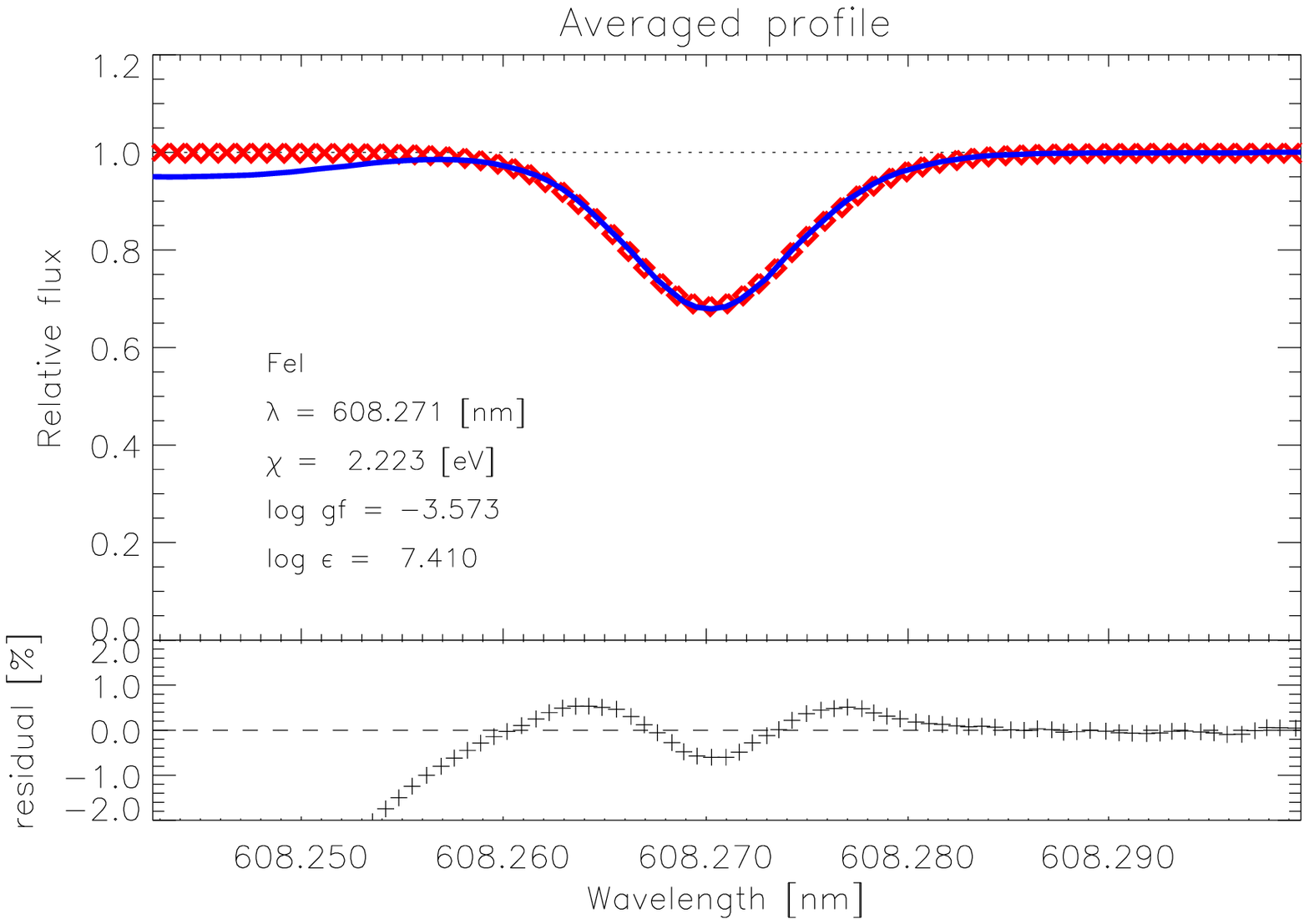}}
\resizebox{\hsize}{!}{\includegraphics{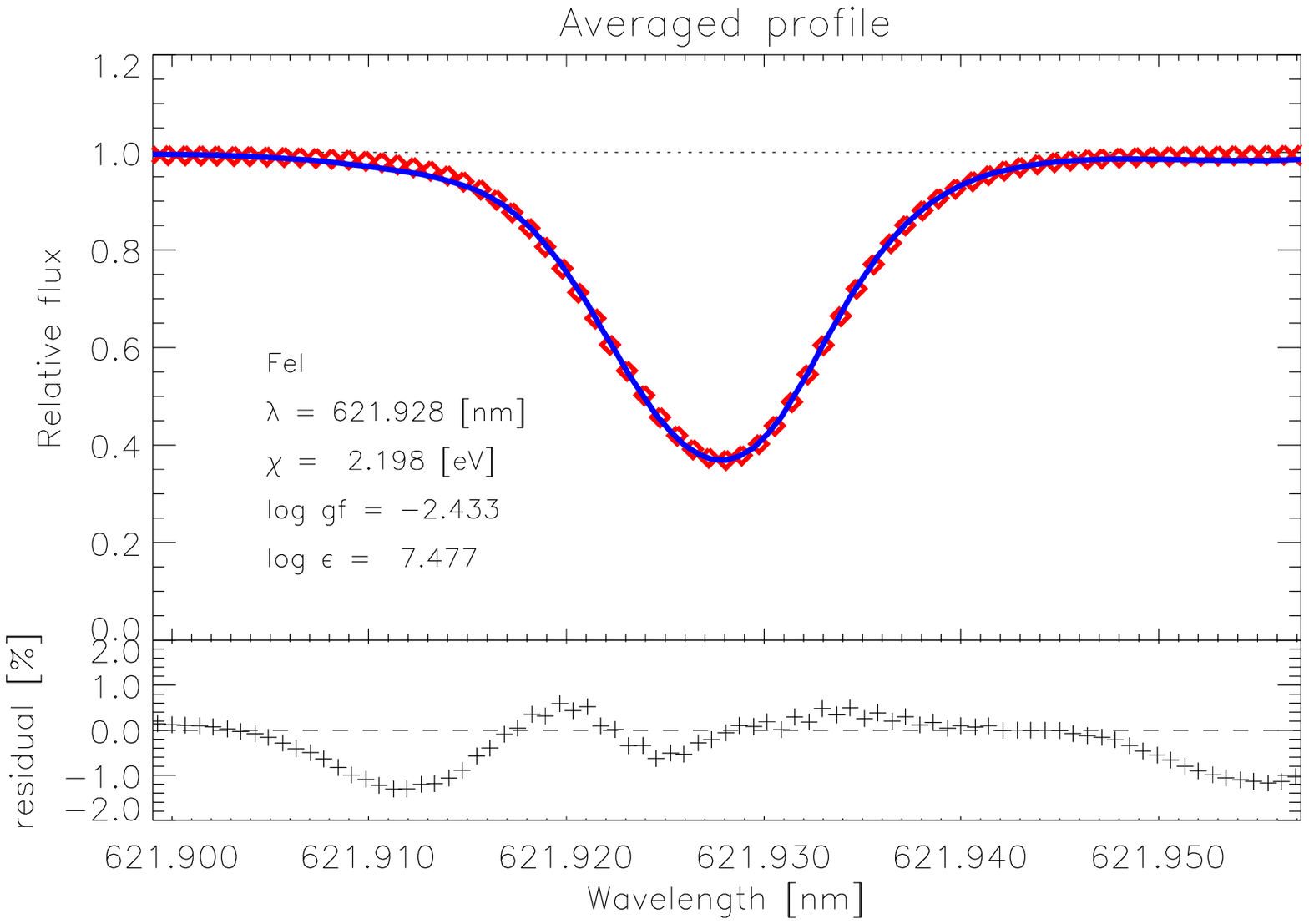}}
%\resizebox{\hsize}{!}{\includegraphics{figures/FeI6082.7_flux_res.ps}}
%\resizebox{\hsize}{!}{\includegraphics{figures/FeI6219.2_flux_res.ps}}
\caption{The predicted flux profiles of the Fe\,{\sc i} 608.2 and 621.9\,nm
lines (diamonds) compared with the solar flux atlas (solid lines,
Kurucz et al. 1984).
The theoretical lines have been disk-integrated using a solar rotational
velocity of $v_{\rm rot} {\rm sin} i = 1.8$\,km\,s$^{-1}$ and convolved
with sinc-function to account for the finite spectral resolution 
of the FTS-atlas}
         \label{f:prof_flux}
\end{figure}

The Doppler shifts introduced by the
convective flow velocities in the photosphere cause significant
line broadening beyond the thermal, radiative and collisional broadening. 
%This is strikingly obvious when solely relying on classical 1D
%model atmospheres which ignores the gas motions: the predicted
%lines are all much too narrow compared with the observed profiles
%(Fig. \ref{f:prof_v=0}). In comparison,
Fig. \ref{f:prof} shows a few examples of spatially and temporally averaged 
Fe\,{\sc i} and Fe\,{\sc ii} lines at disk-center calculated using the 3D solar
simulation together with the corresponding
observed intensity profiles; additional examples are given in Paper II.  
Clearly the agreement is very satisfactory for unblended 
Fe\,{\sc i} and Fe\,{\sc ii} lines with almost a perfect match.
In contrast, theoretical 1D profiles are clearly discrepant in spite
of the presence of both a micro- and macroturbulence (as exemplified
by the Fe\,{\sc i} 621.9\,nm line which is shown 
in the middle panel of Fig. \ref{f:prof}).
The residual intensities amount to only $<1\%$, which is accomplished 
without any micro- or macroturbulent broadening.  
The FWHM of the Fe lines typically agree to within 1\%.
Fig. \ref{f:prof_flux} shows the disk-integrated flux
profiles of two Fe\,{\sc i} lines as compared with the solar flux
atlas of Kurucz et al. (1984) and again the agreement is very encouraging.
In fact, given the good agreement in general for both intensity and
flux profiles, blending lines, which may otherwise have gone undetected 
in a 1D analysis, are easily identified. 
%(as examplified in Paper II and Asplund 2000) 
%as blends distort the profiles and degrade the match. 

\begin{figure}[t]
\resizebox{\hsize}{!}{\includegraphics{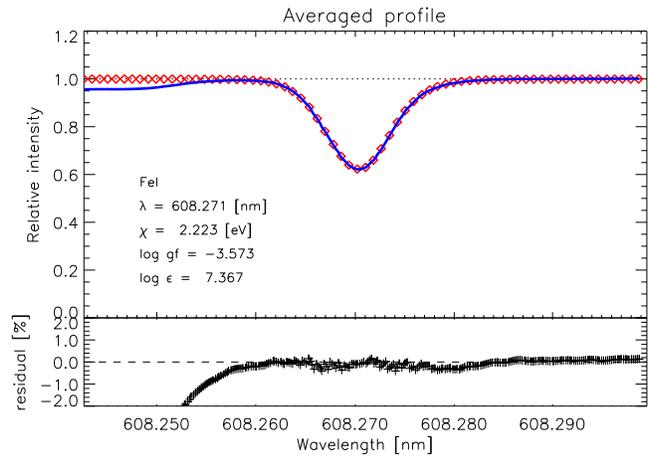}}
%\resizebox{\hsize}{!}{\includegraphics{figures/FeI6082.7_253_res.ps}}
\caption{The predicted spatially and temporally averaged 
profile of the Fe\,{\sc i} 608.2 line (diamonds) using a solar simulation
based on a previous equation-of-state and line opacities 
(Gustafsson et al. 1975 with subsequent updates 
instead of Mihalas et al. 1988 and Kurucz 1993)
compared with the solar disk-center 
intensity atlas (solid lines, Brault \& Neckel 1987). The agreement
is even better than the corresponding profile shown in Fig. \ref{f:prof}
}
         \label{f:prof_gust}
\end{figure}

Although the overall agreement is very satisfactory, 
it is clear from a closer inspection of Figs. \ref{f:prof}
and \ref{f:prof_flux} that there
are systematic discrepancies in the line profiles which are
appearent in most intensity and flux profiles, in particular
the weaker lines. The cores of
the predicted lines tend to be slightly too shallow while the near
line wings are somewhat too broad, which suggests a slightly over-estimated
rms vertical velocity amplitude in the solar simulation. 
The slightly problematic line cores may also signal departures 
from LTE (cf. Rutten \& Kostik 1982), 
which is more likely to affect the cores than the wings.
Furthermore, the cores of
intermediate strong lines tend to be displaced compared with
observations; 
the latter feature will be discussed further in Sects. \ref{s:shift} and
\ref{s:asym}.
It should not come as a surprise that the cores of the stronger
lines show discrepancies, since the highest atmospheric
layers are likely the least realistic due to still missing ingredients in the
simulations and spectral synthesis in terms of e.g. 
departures from LTE and the inexact
line blanketing treatment in the actual convection simulation. 

The minor disagreements shown by the Fe lines are very
important, since they point to how the simulations can be improved further. 
%Rather than hiding the existing problems through a tuning
%of the available free parameters as is unfortunately all-too-often
%done in 1D analyses, one should rather be guided by them since they
%reflect missing ingredients and shortcomings of the modelling.
Prior to the convection simulations used here, we have carried out
several similar solar simulations which differed 
from the present ones, most notably in terms of equation-of-state
and opacities 
(Gustafsson et al. 1975 with subsequent updates vs. Mihalas et al. 1988
and Kurucz 1993),
height extension ($z_{\rm min} = -0.6$ vs. $-1.0$\,Mm) 
and numerical resolution (100\,x\,100\,x\,82, 50\,x\,50\,x\,82 and
50\,x\,50\,x\,63 vs. 200\,x\,200\,x\,82). 
In all cases they suffered from more pronounced problems in terms
of line asymmetries, which were subsequently addressed with
the more refined and improved simulations. 
However, it is noteworthy that the overall shapes of weak Fe lines 
illustrated in Figs. \ref{f:prof} and \ref{f:prof_flux} were even
better described with the previous equation-of-state and line opacities
(Gustafsson et al. 1975 with subsequent updates), 
as seen Fig. \ref{f:prof_gust}, although with a slightly smaller 
Fe abundance. 
One may speculate that the older equation-of-state better describes the
conditions typical of the line-forming layers than the more recent
version (Mihalas et al. 1988) which has been optimized for stellar
interiors.  But it may also be coincidental, since the differences 
in line profiles are relatively small and there are also additional
differences in terms of resolution 
(253\,x\,253\,x\,163) and $z_{\rm min}$ ($-0.6$\,Mm) between the two
simulations. 
A further investigation into the
matter would, however, be interesting. 
%With a poorer numerical resolution on the other hand the 
%predicted lines tend to be too narrow and deep for a given strength
%(Asplund et al. 2000a), since only with
%a sufficient resolution will the tails of the velocity distributions
%be properly sampled (Stein \& Nordlund 1998).  

%%%%%%%%%%%%%%%%%%%%%%%%%%%%%%%%%%%%%%%%%%%%%%%%%%%%%%%%%%%%%%%%%%%%
\section{Solar Fe line shifts \label{s:shift}}
%%%%%%%%%%%%%%%%%%%%%%%%%%%%%%%%%%%%%%%%%%%%%%%%%%%%%%%%%%%%%%%%%%%%

\begin{figure}[t]
\resizebox{\hsize}{!}{\includegraphics{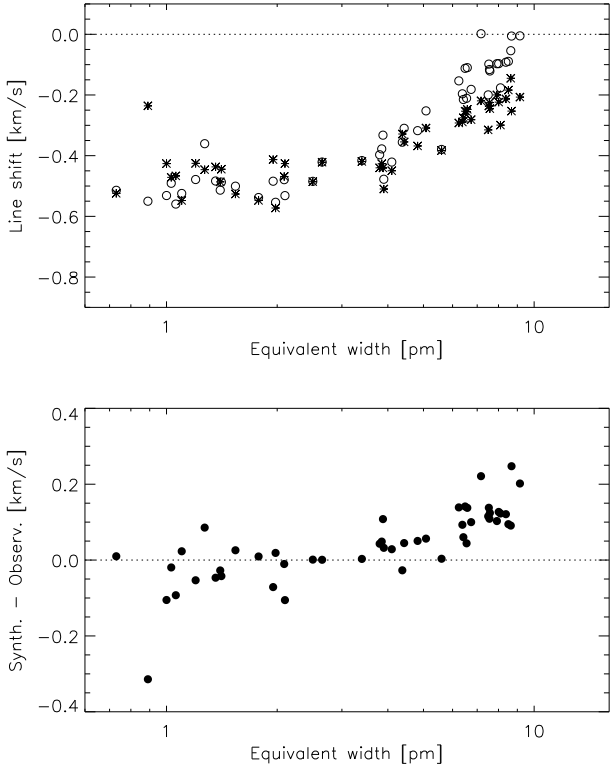}}
%\resizebox{\hsize}{!}{\includegraphics{figures/FeI_shift_201.ps}}
\caption{{\it Upper panel:} The predicted (open circles) 
and observed (stars) line shifts for Fe\,{\sc i}
lines as a function of equivalent widths. The equivalent widths are
taken from Blackwell et al. (1995), Holweger et al. (1995) and 
Moore et al. (1966),
in this order of preference.
%The dashed line correspond to no convective line shift. 
{\it Lower panel:} The differences between predicted and observed
line shifts. The gravitational red-shift of 633\,m\,s$^{-1}$ has
been subtracted from the observed line shifts. 
The agreement is very good for weak and intermediate strong
lines but becomes progressively worse for stronger lines. Most of the
scatter for the weaker lines is probably due to uncertainties in 
laboratory wavelengths and blends
}
         \label{f:feishift}
\end{figure}

\begin{figure}[t]
\resizebox{\hsize}{!}{\includegraphics{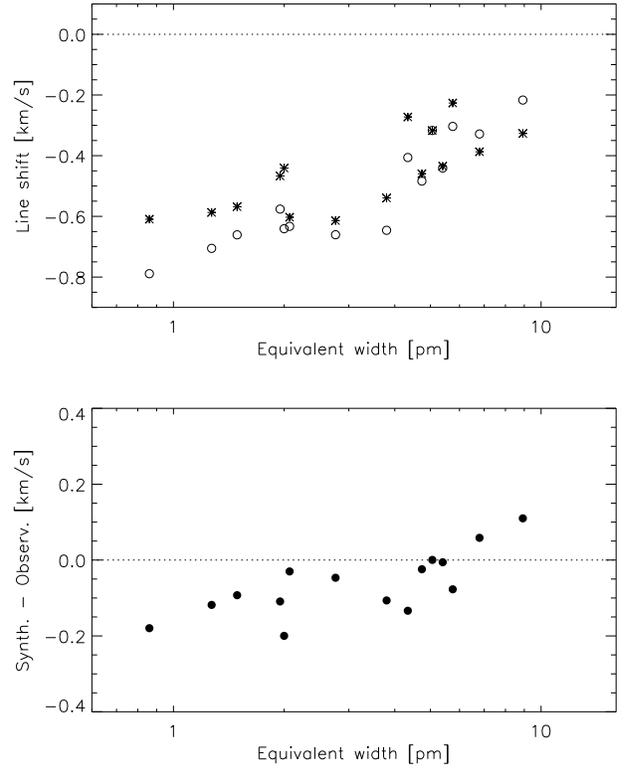}}
%\resizebox{\hsize}{!}{\includegraphics{figures/FeII_shift_201.ps}}
\caption{Same as Fig. \ref{f:feishift} but for the 15 Fe\,{\sc ii} lines.
The equivalent widths are taken from Hannaford et al. (1992)
}
         \label{f:feiishift}
\end{figure}

\begin{figure}[t]
\resizebox{\hsize}{!}{\includegraphics{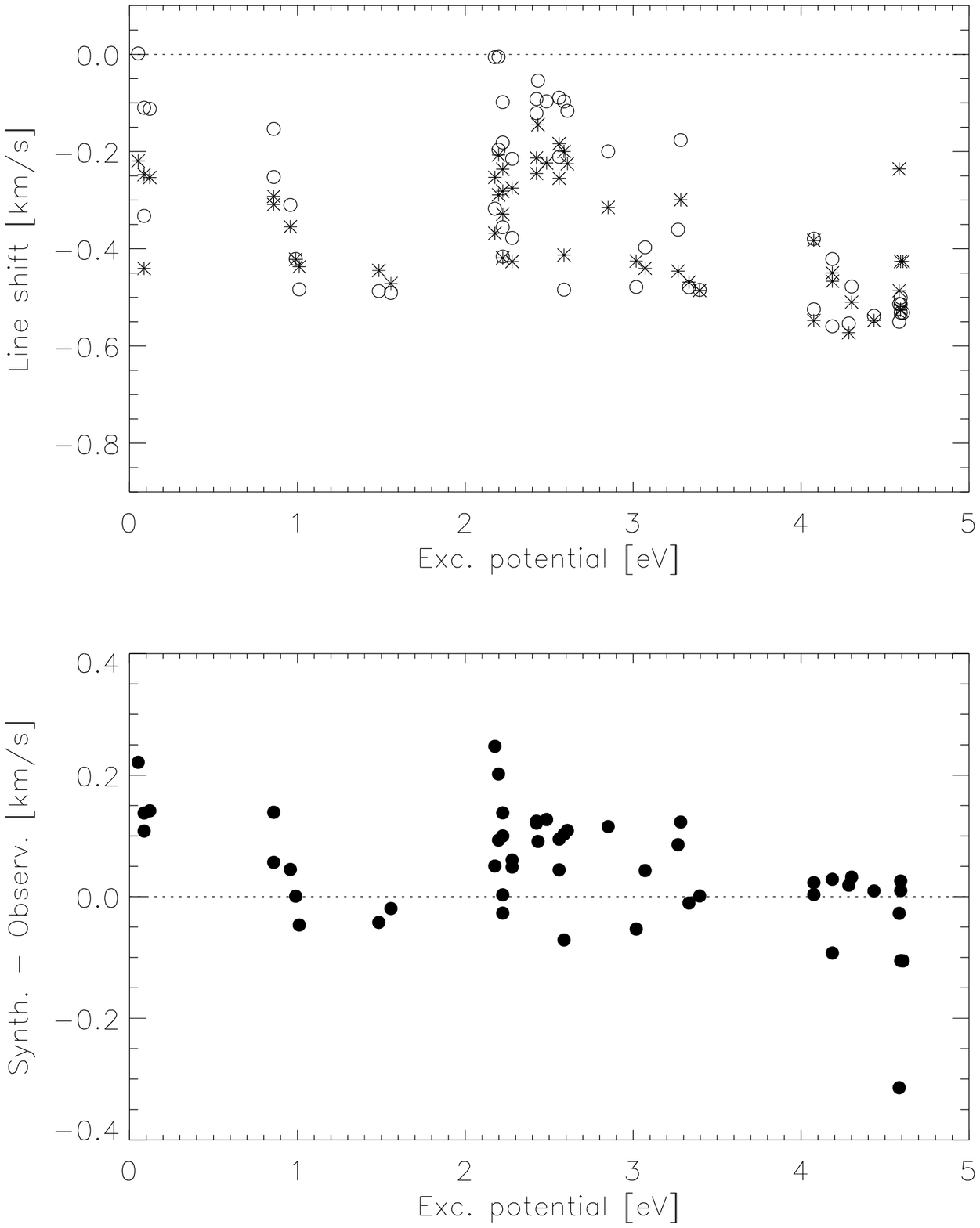}}
%\resizebox{\hsize}{!}{\includegraphics{figures/FeI_shift_chi_201.ps}}
%\centerline{
%\psfig{figure=figures/FeI_shift_chi_201.ps,width=9.cm}}
%  \picplace{1cm}
\caption{Same as Fig. \ref{f:feishift} but
as a function of excitation potential of the lower level. 
In general high excitation lines are formed in deeper layers and
thus have larger convective blueshifts. The trend is not as
striking as in Fig. \ref{f:feishift}, since also e.g. the oscillator
strength determines the line-formation depth
}
         \label{f:feishiftchi}
\end{figure}

A major advantage with solar observations compared with 
corresponding stellar observations is the existence of 
high-quality spectral atlases given on an absolute velocity scale, which
is possible since the differential radial velocity between the Sun and
the Earth can be accurately corrected for 
(e.g. Kurucz et al. 1984; Brault \& Neckel 1987; Neckel 1999). 
Furthermore, the well-determined solar mass and radius
allow the solar gravitational
redshift of 636\,m\,s$^{-1}$, or 633\,m\,s$^{-1}$ for light intercepted on
Earth (Lindegren et al. 1999),
to be estimated, leaving the remaining
shifts to be attributed to convection
as pressure shifts and similar line shifts are of much lesser importance
(e.g. Allende Prieto 1998). 
Solar spectral lines show different
line shifts depending on the typical depths of formation of the
line cores. 
%Since the convective overshoot region occurs in the region
%directly above continuum optical depth unity, lines formed at higher
%layers (stronger lines) will in general show smaller convective blueshifts
%compared with lines formed deeper in in the photosphere. As a result
%There is clear trend between line strength and line shift 
%(Allende Prieto \& Garc\'{\i}a L{\'o}pez 1998a), as well as a
%trend between excitation potential of the lower
%level of the transition and the line shift (Dravins et al. 1981, 1986),
%although the latter is less striking
%since also the transition probability determines the final line strength. 
%The strongest Fe lines, whose cores are formed above the convective
%overshoot region, seem to have no net shift once the gravitational redshift
%is removed (Allende Prieto \& Garc\'{\i}a L{\'o}pez 1998a).
Unfortunately, the remaining uncertainties in the individual laboratory
wavelengths and possible blends cause a scatter of about 
100\,m\,s$^{-1}$ in the observed line shifts, in
particular for the weaker lines. 
%Furthermore, the dependence on e.g.
%wavelength and excitation potential also increases the observed scatter in the 
%$W_\lambda$-line shift diagram (Dravins et al. 1981).

Through the self-consistently calculated convective flows in the
solar simulations the predicted line shifts can be 
directly compared with observations on an absolute wavelength scale.
Figs. \ref{f:feishift} and  \ref{f:feiishift} show 
the calculated line shifts for Fe\,{\sc i} and Fe\,{\sc ii} lines,
respectively; we prefer to plot the shifts vs line strengths rather
than vs line depths as Hamilton \& Lester (1999), since due to saturation
the latter method will not detect clearly the gravitional redshift
plateau (Allende Prieto \& Garc\'{\i}a L{\'o}pez 1998a). 
Weaker lines show as expected more prominent blueshifts
and Fe\,{\sc ii} lines more so than Fe\,{\sc i} lines due to their in
general deeper layers of formation.
The maximum predicted blueshift for Fe\,{\sc i} lines is 550\,m\,s$^{-1}$
while it reaches 800\,m\,s$^{-1}$ for Fe\,{\sc ii} lines.
A trend between line shift and excitation potential is present
(Fig. \ref{f:feishiftchi}) but much less pronounced due to the
obscuration introduced by the $W_\lambda$- and $\lambda$-dependencies
(cf. Fig. \ref{f:bis_abund}, \ref{f:bis_lam} and Hamilton \& Lester 1999).

As clear from Figs. \ref{f:feishift} and  \ref{f:feiishift} the 
predicted line shifts agree well with observations for weak 
and intermediate strong lines.
The average difference for weak 
($W_\lambda \la 6\,$pm) Fe\,{\sc i} lines is 
$0\pm53$\,m\,s$^{-1}$ ($0.000\pm0.053$\,km\,s$^{-1}$), 
while when including also the intermediate
strong lines ($W_\lambda \la 10\,$pm) the corresponding difference 
increases to $51\pm81$\,m\,s$^{-1}$;
in these estimates we have not included the Fe\,{\sc i} 666.8\,nm line
($W_\lambda = 0.89\,$pm) 
as we suspect that it has an erroneous laboratory wavelength, since it
is significantly more discrepant (300\,m\,s$^{-1}$)
than other lines with similar strengths. 
The situation is slightly worse for the Fe\,{\sc ii} lines with a mean
difference of 
$-64\pm85$\,m\,s$^{-1}$ for all 15 lines with $W_\lambda \la 10\,$pm.
However, the laboratory wavelengths for Fe\,{\sc ii} lines
are likely of somewhat poorer quality than the Fe\,{\sc ii} lines
and there are unfortunately only few available lines of the appropriate
strengths. 

\begin{figure}[t]
\resizebox{\hsize}{!}{\includegraphics{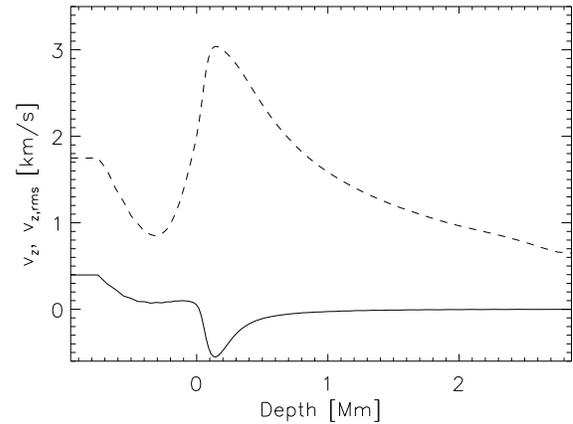}}
%\resizebox{\hsize}{!}{\includegraphics{figures/vz.ps}}
%\centerline{
%\psfig{figure=figures/vz.ps,width=9.cm}}
%  \picplace{1cm}
\caption{The time averaged horizontal average (solid) and 
rms (dashed) vertical velocity in the solar simulations.
Positive vertical velocities indicate downward motion.
Note the upturn in $\langle v_{\rm z} \rangle$ which reaches 
400\,m\,s$^{-1}$ in the uppermost parts of
the atmosphere. The horizontal part of the curves at the
upper boundary is a consequence of the boundary formulation,
which specifies the same vertical velocity in the two top
layers
}
         \label{f:vz}
\end{figure}

\begin{figure}[t]
\resizebox{\hsize}{!}{\includegraphics{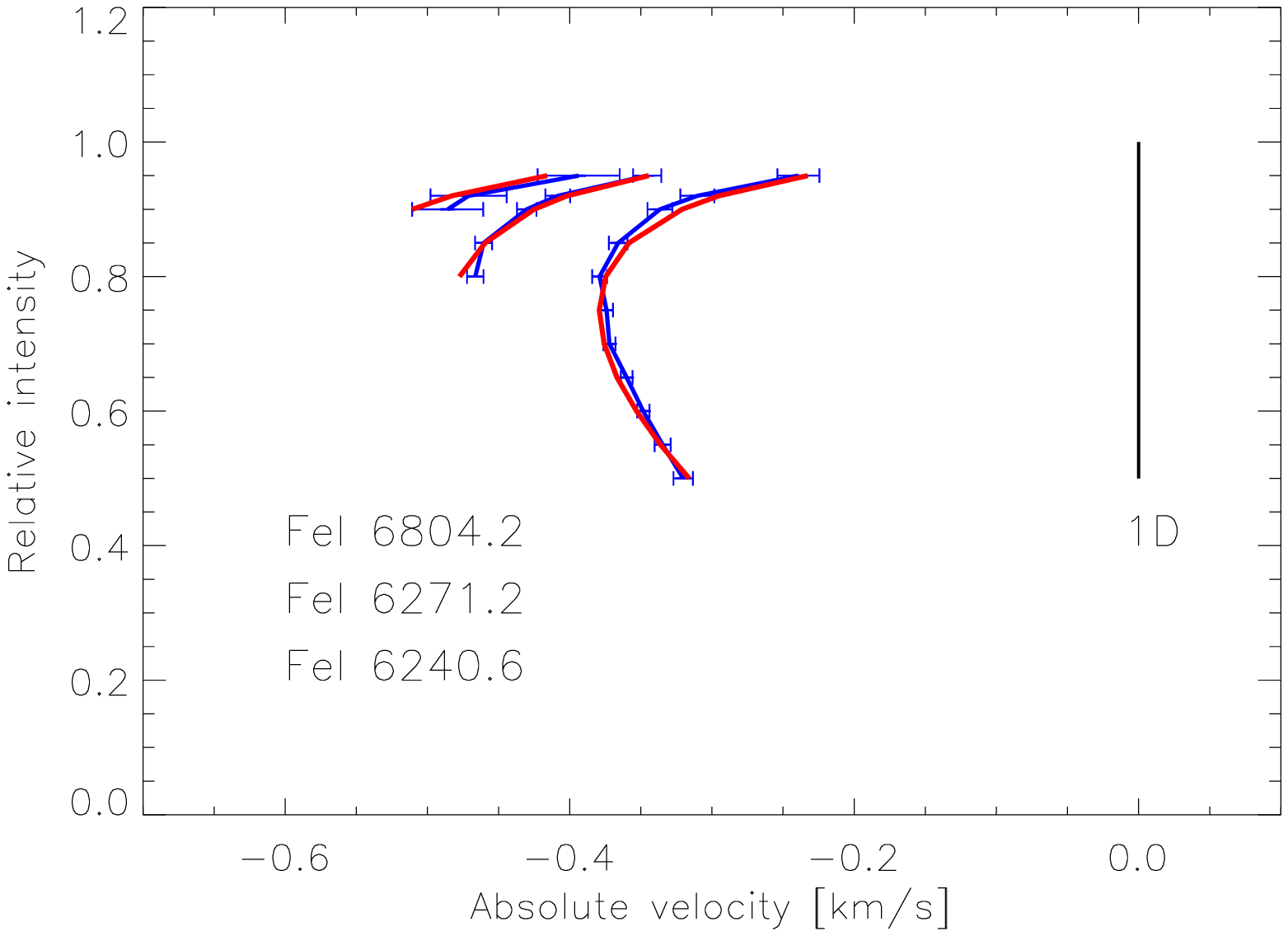}}
%\resizebox{\hsize}{!}{\includegraphics{figures/FeI_bis_201_ex.ps}}
%\centerline{
%\psfig{figure=figures/FeI_bis_201_ex.ps,width=9.cm}}
%  \picplace{1cm}
\caption{The predicted (solid line) and observed (line with error bars)
bisectors of the Fe\,{\sc i} 680.4, 627.1 and 624.0\,nm (in order of increasing 
line strength) lines. It should be emphasized that the velocity scale
is absolute for both the computed and measured line asymmetries, i.e. 
no arbitrary wavelength shifts have been applied in order to bring
the two into agreement. Clearly the predicted bisectors agree
rather satisfactory with the observations. In fact, the close resemblance
for these lines is partly fortuitous since both the laboratory
wavelengths and the wavelength calibration of the solar intensity atlas
may only be accurate to about $30-50$\,m\,s$^{-1}$. In comparison, classical
1D model atmospheres will of course only produce vertical bisectors
at zero absolute velocity}
         \label{f:bis}
\end{figure}

As evident from Figs. \ref{f:feishift} and  \ref{f:feiishift},
the correspondance between predictions and observations 
becomes progressively worse for stronger Fe\,{\sc i} and
Fe\,{\sc ii} lines; a trend with $W_\lambda$ in the line shift 
differences may also extend to weaker Fe\,{\sc ii} lines, according
to Fig. \ref{f:feiishift}. 
We attribute this effect predominantly to the influence of the outer
boundary in the spectral line calculations but departures from LTE
may also play a role. The cores of these 
strong lines have significant optical depths already at the uppermost
depth layers where the average vertical velocity is directed downwards,
as shown in Fig. \ref{f:vz}.
This follows from having on average no net mass flux, since the 
downward moving material in general has smaller densities and
thus larger velocities in these
layers. Since the temperature-velocity correlations present further in has
essentially completely disappeared at these large heights, this results
in convective redshifts, which do not seem to be present
in the observational evidence (Allende Prieto \& Garc\'{\i}a L{\'o}pez 1998a).
The discrepancy was noticably larger in the simulations 
with smaller extension (maximum height of 0.6\,Mm) and numerical resolution
(Asplund et al. 2000a). 
We therefore consider the effect as a
numerical artifact, which should be possible to further reduce by using an 
improved treatment of the outer boundary and higher numerical resolution.   
In spite of the remaining shortcomings, {\it differential} line shifts 
are very accurate; the main 
uncertainty is of observational
(blends and laboratory rest wavelengths) nature.

%%%%%%%%%%%%%%%%%%%%%%%%%%%%%%%%%%%%%%%%%%%%%%%%%%%%%%%%%%%%%%%%%%%%
\section{Solar Fe line asymmetries \label{s:asym}}
%%%%%%%%%%%%%%%%%%%%%%%%%%%%%%%%%%%%%%%%%%%%%%%%%%%%%%%%%%%%%%%%%%%%

\begin{figure}[t]
\resizebox{\hsize}{!}{\includegraphics{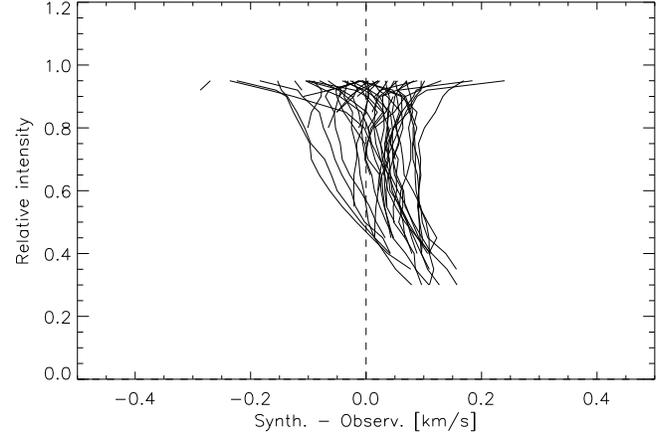}}
%\resizebox{\hsize}{!}{\includegraphics{figures/FeI_bis_201.ps}}
%\centerline{
%\psfig{figure=figures/FeI_bis_201.ps,width=9.cm}}
%  \picplace{1cm}
\caption{Differences between predicted and observed bisectors for
the Fe\,{\sc i} lines used for the present analysis;
a gravitational red-shift of 633\,m\,s$^{-1}$ has been subtracted for
the observed bisectors.
Except for a tendency
for the predicted strong lines (line depths greater than 0.5) 
to have slightly too little blueshift or too much redshift, as also
obvious in Fig. \ref{f:feishift}, 
the agreement is clearly very satisfactory. 
%Given the high level of
%consistency, it is easy to detect problematic lines due to blends
%(a sloping difference between observed and predicted bisectors) and
%erroneous laboratory wavelengths (large velocity offset). 
The line furthest to the left is Fe\,{\sc i} 666.8\,nm, whose laboratory
wavelength therefore can be suspected to be slightly in error 
}
         \label{f:feibis}
\end{figure}

\begin{figure}[t]
\resizebox{\hsize}{!}{\includegraphics{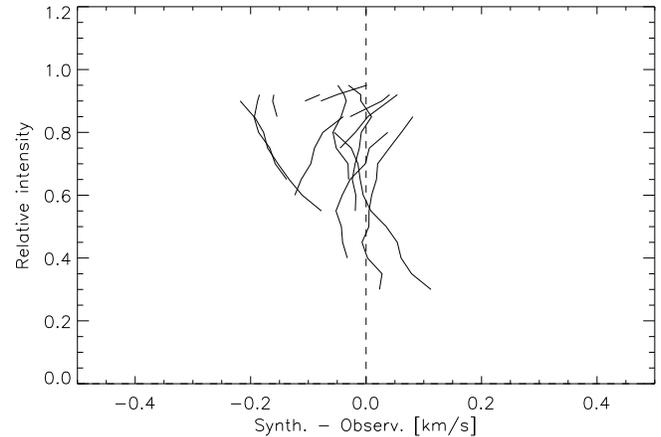}}
%\resizebox{\hsize}{!}{\includegraphics{figures/FeII_bis_201.ps}}
%\centerline{
%\psfig{figure=figures/FeII_bis_201.ps,width=9.cm}}
%  \picplace{1cm}
\caption{Same as Fig. \ref{f:feibis} but for the 15 Fe\,{\sc ii} lines
}
         \label{f:feiibis}
\end{figure}

Line asymmetries carry additional information to line shifts, since
the lines trace the atmospheric structure throughout the line forming
region and not only the higher layers where the core is formed 
and thus the line shift. Fig. \ref{f:bis} shows a few examples of the
predicted line asymmetries and the corresponding observed bisectors.
It should be noted that both types of bisectors are on an absolute
wavelength scale and the synthetic bisectors have therefore not been
shifted in velocity to match the observations.
In order to achieve such a remarkable agreement it is necessary to
have both a very accurate description of the atmospheric structure and the
details of line formation as well as very high quality laboratory 
wavelengths. Clearly the result is very satisfactory. 
The excellent correspondance 
in Fig. \ref{f:bis} is not only fortuitous, as is apparent from
an inspection of Fig. \ref{f:feibis}, which shows the differences
in observed and predicted line asymmetries for all the 67 computed Fe\,{\sc i}
lines. Under ideal conditions the differences should all be vertical
lines with no velocity offset, which in fact is not far from the truth. 
In particular the weaker lines show excellent agreement, while the
situation becomes progressively worse for the cores of the
stronger lines, reflecting
the shortcomings already discussed in Sect. \ref{s:shift}. 
%Due to the influence from the outer boundary the stronger
%lines show too little blueshift or too much redshift.
This discrepancy also affects the bisectors closer to the continuum
for the stronger lines, causing the bisector differences to be
predominantly positive. 
%When only considering weaker lines the full bisectors scatter
%nicely around no net difference, as illustrated in Fig. \ref{f:feibisweak}.
The situation for the Fe\,{\sc ii} lines are shown in Fig. \ref{f:feiibis},
which again is very satisfactory. 
Both with an inferior resolution and height extension in the convection
simulations the resulting bisectors are of noticably lower quality 
when comparing with observations (Asplund et al. 2000a).
The good overall agreement in terms of line asymmetries therefore
lend very strong support to the realism of the convection simulations.
%and suggests that the necessary numerical resolution for
%spectroscopic purposes has already been achieved.

In fact it is very easy to detect problematic lines due to erroneous
laboratory wavelengths (large velocity offset) and blends 
(discrepant bisector shape) when comparing line asymmetries.
%According to Fig. \ref{f:feibis}, the laboratory wavelength of the 
%Fe\,{\sc i} 666.8\,nm line may be questioned,
%since it is offset by about 300\,m\,s$^{-1}$.
%Likewise, we suspect that the Fe\,{\sc ii} 643.2 and 651.6\,nm lines are
%blended since they show unusually pronounced slopes in Fig. \ref{f:feibis}.
It can be mentioned that initially the wavelengths for the
Fe\,{\sc i} 697.2 and 718.9\,nm lines were accidentally taken from
the VALD database rather than from Nave et al. (1994),
which differed by only 16 and 5\,m\AA $ $ (corresponding to
0.7 and 0.2\,km\,s$^{-1}$), respectively, though
immediately detected when analysing the line asymmetries.
Furthermore, the Fe\,{\sc ii} wavelengths provided
by Johansson (1998, private communication)
were found to be of significantly higher quality than the ones 
given in Hannaford et al. (1992).

%\input{old2d3d}

%%%%%%%%%%%%%%%%%%%%%%%%%%%%%%%%%%%%%%%%%%%%%%%%%%%%%%%%%%%%%%%%%%%%
\section{The nature of macro- and microturbulence}
%%%%%%%%%%%%%%%%%%%%%%%%%%%%%%%%%%%%%%%%%%%%%%%%%%%%%%%%%%%%%%%%%%%%

The concepts of micro- and macroturbulence are introduced in
1D analyses in order to account for the missing line broadening
on length scales less than and larger than a unit optical depth,
respectively. Naturally such a simplistic division is artificial
since the motions occur on a range of scales. Furthermore, they are
supposed to represent turbulent motions and thus are normally
assumed to be isotropic. In reality the appearance of the
photospheric granular velocities is clearly more laminar 
than turbulent with distinct upflows and downflows, a direct consequence
of the strong density stratification (Nordlund et al. 1997).
Given the evidences presented here and in Papers II and III, there
appears to be no need to invoke any macro- or microturbulence
in spectral syntheses based on realistic 3D model atmospheres. 

The excellent agreement between predicted and observed line profiles
shown in Fig. \ref{f:prof} implies that the theoretical lines have
the correct widths without the use of any macroturbulence. The 
classical concept of macroturbulence can therefore be fully explained by
the self-consistently calculated convective velocity fields and the
stellar oscillations. It is less obvious that the simulations properly
account for the small-scale motions normally referred to as 
microturbulence given the finite numerical resolution which may or may not
resolve all significant velocities. We believe, however, that the
currently best solar simulations described here are
of sufficiently high resolution to describe also the most important 
effects of these small-scale motions.
Firstly, solar simulations with different numerical grid resolutions 
indicate that the velocity distributions have essentially 
converged already at a resolution of 200\,x\,200\,x\,82, which
is also apparent from the predicted line shapes (Asplund et al. 2000a).
The contribution from unresolved scales should therefore be very minor.
We emphasize that a high resolution is needed in the construction of
the 3D model atmospheres to allow the high-velocity tails of the
velocity distributions, but that smaller resolutions can be used for
the spectral synthesis, 
as also verified by extensive testing of the line formation in model atmospheres
interpolated to various resolutions from the original 
200\,x\,200\,x\,82 data cubes prior to the analysis presented here.
Secondly, the derived Fe\,{\sc ii} and Si abundances show no
trend with line strength when using profile fitting (Papers II and III), which
suggests that the velocities important for the line broadening have
already been accounted for without resorting to the use of extra
microturbulent-like velocities. 
There is, however, a minor trend for Fe\,{\sc i} lines (Paper II) but
since  Fe\,{\sc ii} and Si lines should be affected similarly yet show
no dependence with line strength, we attribute the problem with 
the Fe\,{\sc i} lines rather to signatures of departures from LTE. 
Fe\,{\sc i} lines should be more susceptible to NLTE effects while 
Fe\,{\sc ii} lines are essentially immune to such effects for solar-type
stars (Shchukina \& Trujillo Bueno 2000). From the results presented here
and in Papers II and III there is no indication that any extra
microturbulent broadening must be included in a fashion similar to
that of Atroshchenko \& Gadun (1994). We attribute this difference mainly to
our use of much higher resolution (their solar simulations had
only $30^3$ or $32^3$ grid-points),
height extension (their spectral line calculations only extend 
up to about 400\,km above $\tau_{\rm Ross} \simeq 1$)
and temporal coverage (their calculations were restricted to only
one or two snapshots), which allow our
simulations to better describe the full effects of the convective motions.
Therefore, either from a pragmatic point of view or by advocating 
the principle of Occam's razor, both macro- and microturbulence appear
redundant in 3D analyses.

Thus the predominant explanation is the same for micro- and macroturbulence,
namely the photospheric granular velocity field and temperature inhomogeneities.
Additionally, photospheric oscillatory motions play a role in the
overall macroturbulent broadening of the lines without 
affecting the line strengths. The main component for
the microturbulence therefore
does not at all arise from the microscopic turbulent motions 
%and energy cascades sometimes associated with a high Reynolds-number convection
but rather from (gradients in) convective motions that are resolved with
the current simulations.
The small scale energy cascades and turbulence 
associated with a high Reynolds-number plasma 
are present in the solar convection zone 
but their {\em intensities} are very small
in the upflows to which the line strengths are strongly biased 
(Sect. \ref{s:prof_xy}) and therefore they do not influence the line formation
significantly.

%%%%%%%%%%%%%%%%%%%%%%%%%%%%%%%%%%%%%%%%%%%%%%%%%%%%%%%%%%%%%%%%%%%%
\section{Concluding remarks}
%%%%%%%%%%%%%%%%%%%%%%%%%%%%%%%%%%%%%%%%%%%%%%%%%%%%%%%%%%%%%%%%%%%%

The good agreement with
observed line shapes, shifts and asymmetries, lend very strong support
to the realism of the 3D convection simulations. No doubt the
3D predictions are superior to those obtained
from 1D analyses. 
In particular the use of mixing length parameters,
equivalent widths, macro- and microturbulence
no longer appear to be needed.
Therefore derived results, such as elemental abundances, should
be more reliable. 
In spite of the minor remaining shortcomings, 
the overall significant accomplishment is therefore still obvious: 
{\it starting from
only the well-known radiative-hydrodynamical conservation equations
and with no adjustable free parameters besides the treatment
of the numerical viscosity  
in the construction of the 3D model atmospheres, 
detailed line profiles and asymmetries can be predicted which agree 
almost perfectly with observations
and furthermore are far superior to classical 1D predictions with 
several tunable parameters.} It should be stressed, however, 
that the numerical viscosity is merely introduced for
numerical stability purposes and is determined from standard hydrodynamical
test cases with no adjustments allowed to improve the agreement with
observations. In this respect the viscosity is not a freely
adjustable parameter like e.g. the various mixing length parameters
in 1D models.

It is important to emphasize that this accomplishment is only
possible if the convection simulations are highly realistic,
both in terms of input physics (equation-of-state, opacities etc)
and numerical details (numerical and physical resolution, extension,
boundary conditions, radiative transfer treatment etc).
From our various experiments and test calculations we can conclude
that of special importance is the dimension (2D is not adequate for
spectral synthesis, Asplund et al. 2000a), 
resolution (even $\simeq 100^3$ is not
quite sufficient for detailed line shapes, Asplund et al. 2000a) 
and height extension (to limit the influence from the outer boundary) 
of the numerical box. 
Furthermore, in order to achieve the correct temperature structure
it is important to include the effects of line-blanketing in the
3D radiative transfer during the simulations (Stein \& Nordlund 1998). 
We believe that we have now addressed all of these specific issues with
the new generation of convection simulations, which is supported by
the very close resemblance with observed line 
profiles, shifts and asymmetries, as presented here.

All of the above-mentioned features are currently affordable with present-day
supercomputers. 
The obvious disadvantage with this 3D procedure is of course the
time-consuming task to perform the necessary 3D convection simulations and
spectral line calculations, even with the simplifying assumptions of 
opacity binning and LTE.
Furthermore, still only a relatively 
small part of the Hertzsprung-Russell diagram
of solar-type stars has been explored.  
The situation will, however, improve significantly 
during the coming years due to faster computers and more 
efficient numerical algorithms.
Of particular interest will be to extend
the modeling to additional metal-poor stars (cf. Asplund et al. 1999),
A-F type stars and red giants, 
where the granulation is expected to be much more vigorous
than for the Sun, which should therefore influence the line formation more.

%%%%%%%%%%%%%%%%%%%%%%%%%%%%%%%%%%%%%%%%%%%%%%%%%%%%%%%%%%%%%%%%%%%%
\begin{acknowledgements}
We are grateful to Sveneric Johansson for providing us with
unpublished Fe\,{\sc ii} laboratory
wavelengths and information regarding Fe wavelength measurements.
We thank Paul Barklem for providing us with unpublished collisional
broadening data for Fe\,{\sc i} lines and helpful discussions.
We gratefully acknowledge Robert L. Kurucz for the use of his unpublished, 
low Fe opacity distribution functions. 
The constructive suggestions by an anonymous referee are much appreciated.
MA and CAP acknowledge the generous financial support of Nordita.
The convection simulations were performed at UNI$\bullet$C
supercomputing center in Denmark, whose support is greatly appreciated. 
Extensive use have been made of the VALD database
(Kupka et al. 1999), which has been very helpful.
\end{acknowledgements}

%%%%%%%%%%%%%%%%%%%%%%%%%%%%%%%%%%%%%%%%%%%%%%%%%%%%%%%%%%%%%%%%%%%%

\end{document}